\documentclass[fleqn,usenatbib]{mnras}

\usepackage{newtxtext,newtxmath}

\usepackage[T1]{fontenc}

\DeclareRobustCommand{\VAN}[3]{#2}
\let\VANthebibliography\thebibliography
\def\thebibliography{\DeclareRobustCommand{\VAN}[3]{##3}\VANthebibliography}

\usepackage{graphicx}	

\usepackage[dvipsnames]{xcolor}
\usepackage{enumitem}
\usepackage{ulem}

\newcommand{\removed}[1]{\relax}
\newcommand{\inserted}[1]{#1}

\title[Carbon stars as standard candles]{Carbon stars as standard candles: II. The median $J$ magnitude as a distance indicator}

\author[Parada et al.]{
Javiera Parada,$^{1}$\thanks{E-mail: jparada@phas.ubc.ca}
Jeremy Heyl,$^{1}$
Harvey Richer,$^{1}$
Paul Ripoche$^{1}$
\newauthor{ and Laurie Rousseau-Nepton$^{2,3}$}
\\
$^{1}$Department of Physics and Astronomy, University of British Columbia, 6224 Agricultural Road, Vancouver, British Columbia V6T 1Z4, Canada\\
$^{2}$ Canada-France-Hawaii Telescope, Kamuela, HI, 96743, USA\\
$^{3}$ Department of Physics and Astronomy, University of Hawaii at Hilo, Hilo, HI, 96720, USA
}

\date{Accepted 2020 November 21. Received 2020 November 17; in original form 2020 June 24}

\pubyear{2020}

\begin{document}
\label{firstpage}
\pagerange{\pageref{firstpage}--\pageref{lastpage}}
\maketitle

\begin{abstract}
We introduce a new distance determination method using carbon-rich asymptotic giant branch stars (CS) as standard candles and the Large and Small Magellanic Clouds (LMC and SMC) as the fundamental calibrators. We select the samples of CS from the ($(J-K_{s})_0$, $J_0$) colour-magnitude diagrams, as, in this combination of filters, CS are bright and easy to identify. We fit the CS $J$-band luminosity functions using a Lorentzian distribution modified to allow the distribution to be asymmetric. We use the parameters of the best-fit distribution to determine if the CS luminosity function of a given galaxy resembles that of the LMC or SMC. Based on this resemblance, we use either the LMC or SMC as the calibrator and estimate the distance to the given galaxy using the median $J$ magnitude ($\overline{J}$) of the CS samples. We apply this new method to the two Local Group galaxies NGC 6822 and IC 1613. We find that NGC 6822 has an "LMC-like" CS luminosity function while IC 1613 is more "SMC-like". Using the values for the median absolute $J$ magnitude for the LMC and SMC found in Paper I we find a distance modulus of $\mu_{0}=23.54\pm0.03$ (stat) for NGC 6822 and $\mu_{0}=24.34\pm0.05$ (stat) for IC 1613.


\end{abstract}




\section{Introduction}
\label{sec:intro}

On short-distance scales the determination of $H_0$ depends sensitively on the accuracy of the distances measured to galaxies in the local universe. A number of distance indicators are being studied in order to improve the local observational determination of $H_0$, e.g. the tip of the red giant branch (TRGB) \citep{beaton_2016, 2019ApJ...882...34F}, Cepheids \citep{2016ApJ...826...56R, 2019ApJ...876...85R} and Mira variables \citep{2013IAUS..289..101C, huang_2018}.

Another distance indicator that has recently resurfaced are carbon-rich asymptotic giant stars (CS) \citep{paper1, 2020ApJ...899...66M, 2020ApJ...899...67F}. CS are produced when asymptotic giant branch (AGB) stars experience sufficient dredge-up to bring enough carbon from their core to the surface to make the carbon to oxygen ratio greater than one. This occurs near the tip of the AGB when the star is more luminous than any other red giants. The distinct spectral features of CS make them appear redder than regular AGB stars -- especially in the near-infrared. Their high luminosity coupled with their significantly red colour makes them easily distinguishable from the rest of the stars in the colour magnitude diagram (CMD) making CS very useful distance indicators.

\cite{richer_1984} were the first to use photometric data for CS to obtain the distance modulus of another galaxy (NGC 205). \citeauthor{richer_1984} based their study on the assumption that the CS of NGC 205 have the same mean magnitude as the CS in the Large and Small Magellanic Clouds (LMC and SMC), and thus that these stars could be used as calibrators. This assumption was studied by \cite{2001ApJ...548..712W}, who came to the conclusion that given the small $(J-K_{s})$ colour range for CS, the magnitude spread will also be small. 

In this paper we develop a method to estimate the distances to Magellanic type galaxies using CS as a standard candle and the Magellanic Clouds as our fundamental calibrator. In \citet{paper1} (from now on Paper I) we showed how we determined the colour selection of the CS in the near-infrared colour magnitude diagram. We also analyzed the luminosity functions of these stars in the LMC and SMC finding slight differences between their median absolute $J$ magnitudes. The difference was attributed to the lower metallicity of the SMC as compared to the LMC (see Sec. \ref{sec:ZandSFH} of this paper for details and literature review). The differences in the luminosity functions led us to the realization that using both the Magellanic Clouds as calibrators could compensate for these differences. We will establish how we determine the best calibrator to be used in a case by case basis. 


\section{Data Analysis \& Catalogues}
\label{sec:data}

\subsection{Magellanic Clouds}
\label{subsec:mcdata}

We obtained two separate catalogues for the LMC and SMC from the Two Micron All Sky Survey (2MASS) \footnote{\url{https://irsa.ipac.caltech.edu/Missions/2mass.html}} \citep{2006AJ....131.1163S}. 2MASS uniformly scanned the entire sky in three near-infrared bands; $J$ (1.235 $\mu m$), $H$ (1.662 $\mu m$), and $K_{s}$ (2.159 $\mu m$) using two 1.3 m telescopes located at Mt. Hopkins, AZ, and CTIO, Chile. 

The 2MASS catalogues have a high contamination from foreground Milky Way stars. To remove the foreground contamination, we cross-matched the 2MASS stars with those found in {\it Gaia} DR2 \citep{GaiaMission, GaiaDR2}. We then used proper motions to select stars considered to be LMC and SMC members. 

The 2MASS 99\% completeness magnitude limits are $J=$15.8, $H=$15.1 and $K_{s}=$14.3. The stars in both the LMC and SMC that we will use to calculate distances to other galaxies, are brighter than this limit. Therefore, incompleteness is not an issue and we do not need to correct our catalogues for it.

As we will use the LMC and SMC as calibrators, we require the catalogues in absolute magnitudes. For this, we correct the apparent magnitudes for extinction and reddening, and include the true distance moduli to these galaxies. For the LMC we adopt the mean distance modulus determined by \cite{2019Natur.567..200P}: $\mu_{\mathrm{LMC}}$ = 18.477 $\pm$ 0.004 (statistical) $\pm$ 0.026 (systematic). For the SMC we take the distance modulus found by \cite{2016ApJ...816...49S}: $\mu_{\mathrm{SMC}}$ = 18.96 $\pm$ 0.03 (statistical) $\pm$ 0.05 (systematic). To correct for reddening and extinction we use the reddening maps for both the Magellanic Clouds obtained by \citet{2020ApJ...889..179G} and the extinction coefficients for the 2MASS bands from \cite{2003ApJ...594..279G}.

\begin{figure}
\centering
\includegraphics[width=\columnwidth]{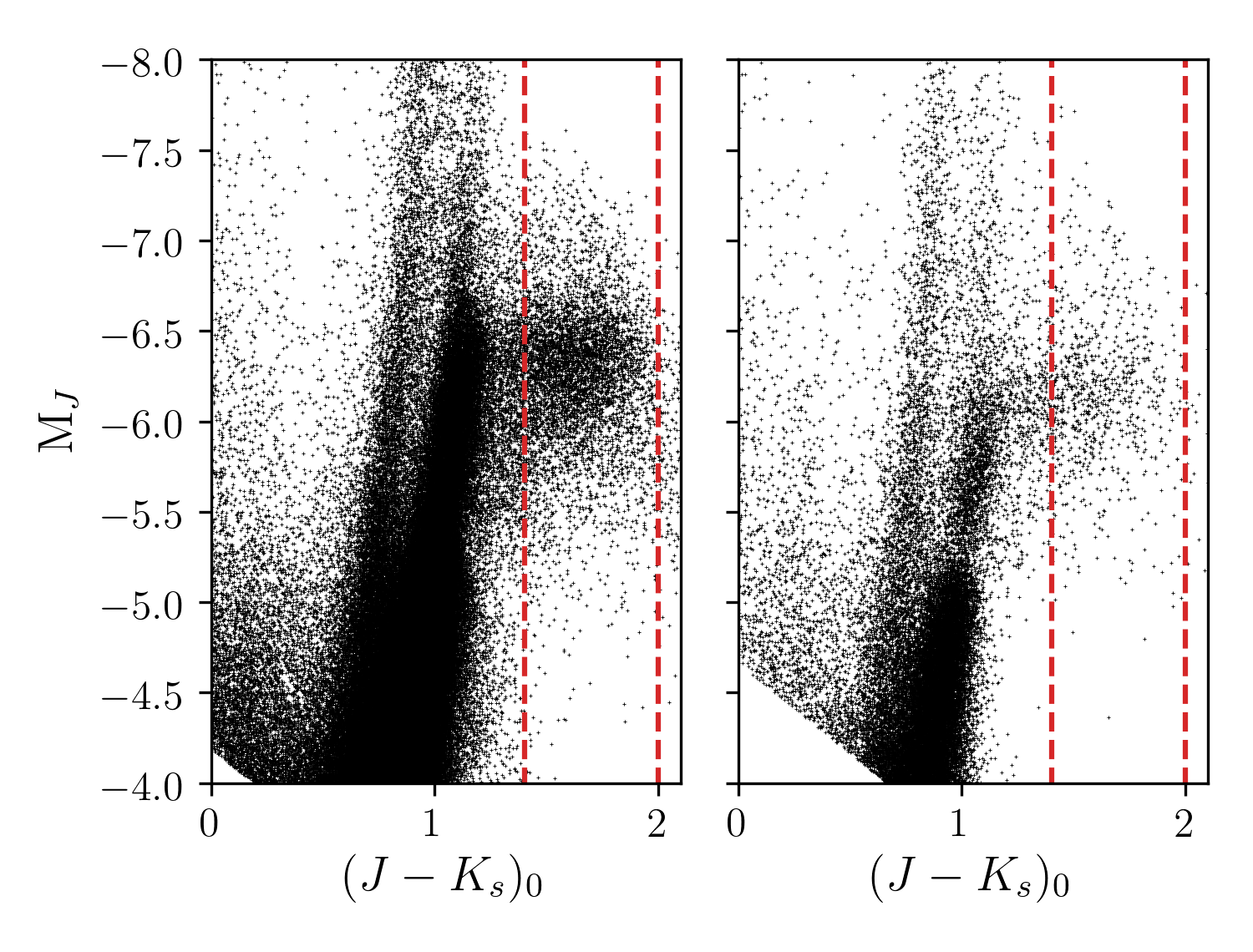}
\caption{$(J-K_{s}$)$_{0}$,M$_{J}$ colour-magnitude diagrams for the Large (\textit{left}) and Small (\textit{right}) Magellanic Clouds. The data shown are the resulting catalogues after filtering foreground contamination and correcting for reddening. The vertical red lines represent the colour limits for the carbon star region (1.4$<$ $(J-K_{s})_{0}$ $<$ 2.0).}
\label{fig:CSselMCs}
\end{figure}

The resulting foreground cleaned $(J-K_{s})_{0}$,M$_{J}$ CMD for the LMC and SMC are shown in Fig. \ref{fig:CSselMCs}. A detailed description of how we constructed the LMC and SMC catalogues used in this paper can be found in Paper I. 

\subsection{NGC 6822}
\label{subsec:ngc6822phot}

\subsubsection{Observations \& Photometry}
The data for NGC 6822 were obtained from observations made with the Wide-field InfraRed Camera (WIRCam) at the Canada-France-Hawaii Telescope (CFHT) in June 2016 and May 2019 (program ID 17AD82, P.I. Laurie Rousseau-Nepton).  The WIRCam observations covered a field of $\sim20$ arcmin$^2$ centred near to the centre of the galaxy. Individual images were taken with three different filters: $J$ (1.253 $\mu$m), $H$ (1.631 $\mu$m), and $K_{s}$ (2.146 $\mu$m). The median seeing values during the observations were 0.58 arcsec for $J$, 0.53 arcsec for $H$ and 0.56 arcsec for $K_{s}$. A total of 40, 160 and 120 images were obtained for $J$, $H$ and $K_{s}$ respectively, giving combined exposure times of 2400 seconds for $J$ and $H$, and 3000 seconds for $K_{s}$.
For each individual image, sources were initially identified using SExtractor \citep{1996A&AS..117..393B}. The SExtractor catalogues were then processed by SCAMP \citep{2006ASPC..351..112B} to find the astrometric solution between the images in the same filter.  Finally, the astrometic solution was used by SWarp \citep{2002ASPC..281..228B} to produce a co-added image for each filter. We performed PSF photometry on the co-added images using DAOPHOT II \citep{daophot} and ALLSTAR \citep{allstar}. The final catalogues were aligned and combined into a single master catalogue. 

To calibrate the instrumental magnitudes obtained from DAOPHOT we used the magnitude difference between the WIRCam photometry and 2MASS stars. Before calculating the difference in magnitude we had to consider that the WIRCam $JHK_{s}$ filters are not identical to those used by 2MASS. Therefore we transformed our photometry to the 2MASS system using the equations provided by the WIRwolf image stacking pipeline\footnote{\url{http://www.cadc-ccda.hia-iha.nrc-cnrc.gc.ca/en/wirwolf/docs/filt.html}}:
\begin{equation}
\begin{split}
&J_{\mathrm{2MASS}} = J_{\mathrm{WIRCam}} + 0.071 \times (J-H)_{\mathrm{WIRCam}} \\
&H_{\mathrm{2MASS}} = H_{\mathrm{WIRCam}} - 0.034 \times (J-H)_{\mathrm{WIRCam}} \\
&K_{\mathrm{2MASS}} = K_{\mathrm{WIRCam}} - 0.062 \times (H-K)_{\mathrm{WIRCam}}.\\
& \quad \quad \qquad +0.002 \times (J-H)_{\mathrm{WIRCam}}.
\end{split}
\end{equation}
We then matched the stars in our master catalogue to 2MASS stars in the same region of the sky as NGC 6822, using the CDS cross-match service\footnote{\url{http://cdsxmatch.u-strasbg.fr/}}. Considering only the stars whose positions matched within 0.15 arseconds, for each filter, we calculated the mean magnitude difference between the stars from our photometry and the 2MASS stars. 

\begin{figure}
\centering
\includegraphics[width=\columnwidth]{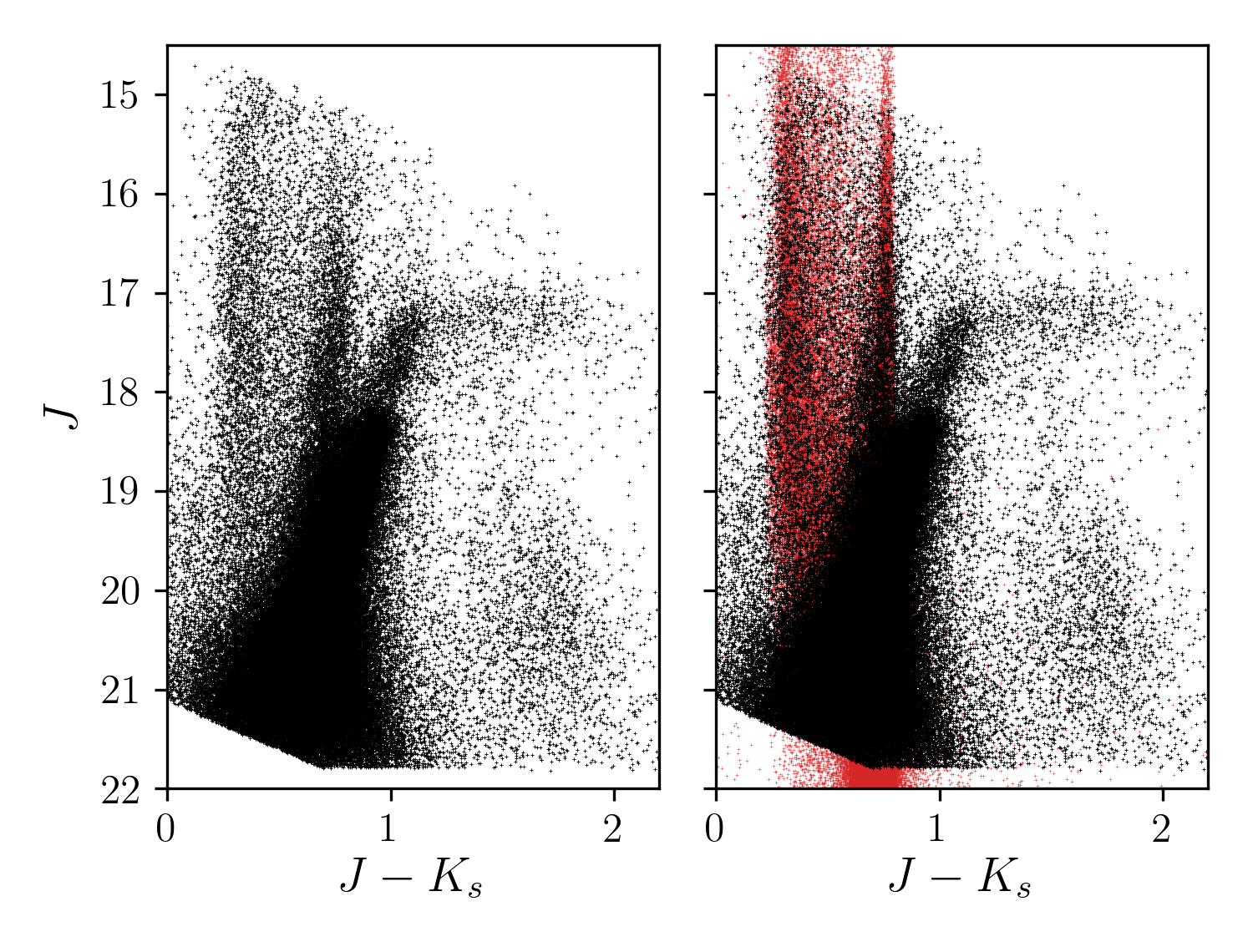}
\caption{\textit{Left:} Calibrated and reddening corrected ($(J-K_{s})_{0}$,$J_0$) colour-magnitude diagram for NGC 6822. \textit{Right:} Same as \textit{left} panel plotted over a synthetic population of foreground Milky Way stars (red) generated with Trilegal  \citep{2005A&A...436..895G}.}
\label{fig:ngc6822cmd}
\end{figure}

NGC 6822 is located at a galactic latitude of about $-18^\circ$, which results in a considerable amount of foreground stellar contamination as well as significant reddening. According to \cite{2011ApJ...737..103S} the Milky Way reddening value along the line-of-sight towards NGC 6822 is $E(B-V$) = 0.21. Other studies using Cepheids \citep{2006ApJ...647.1056G, 2012A&A...548A.129F} and colour-magnitude diagrams  \citep{2014ApJ...794..107R} estimate the total reddening in the line-of-sight to be about $E(B-V)\sim 0.35$ mag.
We adopt the latter value, $E(B-V)=0.35$, to deredden our data as this value includes both the foreground reddening and an estimate of the internal reddening within the galaxy. The calibrated and reddening corrected colour-magnitude diagram is shown in the left panel of Fig. \ref{fig:ngc6822cmd}. 

\subsubsection{Colour-magnitude diagram cleaning}

The right panel of Fig. \ref{fig:ngc6822cmd} displays the NGC 6822 CMD plotted over a theoretical model distribution of Milky Way stars obtained from the population synthesis code Trilegal\footnote{\url{http://stev.oapd.inaf.it/cgi-bin/trilegal}} \citep{2005A&A...436..895G}. Both the modeled foreground stars and our photometry have been dereddened.  Even though there are a large number of foreground stars, these are bluer than the CS blue limit ($(J-K_{s})_{0}=1.4$). Therefore, we do not correct for the presence of foreground stars in the colour-magnitude diagram.

The CMD of NGC 6822 also suffers from a considerable amount of contamination from background galaxies. Despite the fact that these sources are generally fainter than the bulk of the CS, there is a large number of them located in the same colour range ($1.4<(J-K_{s})_{0}<2.0$). Unlike stars, galaxies are extended sources and we can distinguish them from the rest of the point-like stellar sources. In the final step of the PSF photometry, ALLSTAR calculates a sharpness (SHARP) and goodness-of-fit (CHI) index for each star. SHARP and CHI are obtained by comparing different properties of the observed sources to those of the PSF. Sources with SHARP and CHI values well above the mean of the SHARP and CHI value distributions, are generally background galaxies or blends and can be eliminated. 

\begin{figure}
\centering
\includegraphics[width=\columnwidth]{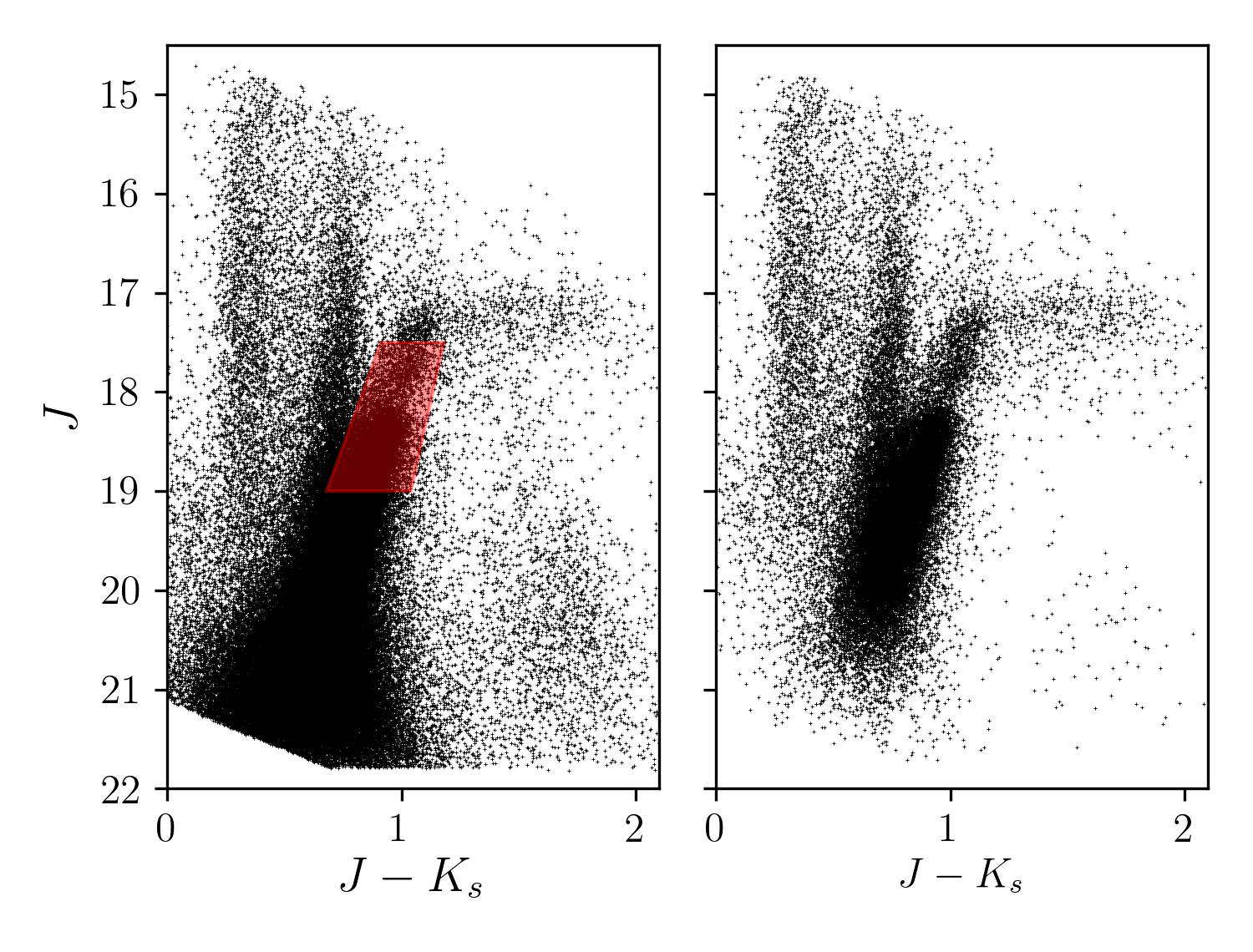}
\caption{\textit{Left:} Same as Fig. \ref{fig:ngc6822cmd} with red box indicating the selected red giant stars used for SHARP and CHI cutoffs. \textit{Right:} The colour-magnitude diagram after applying the SHARP and CHI cutoffs depicted in Fig. \ref{fig:ngc6822sharpchi}.}
\label{fig:ngc6822clean}
\end{figure}

\begin{figure}
\centering
\includegraphics[width=7cm]{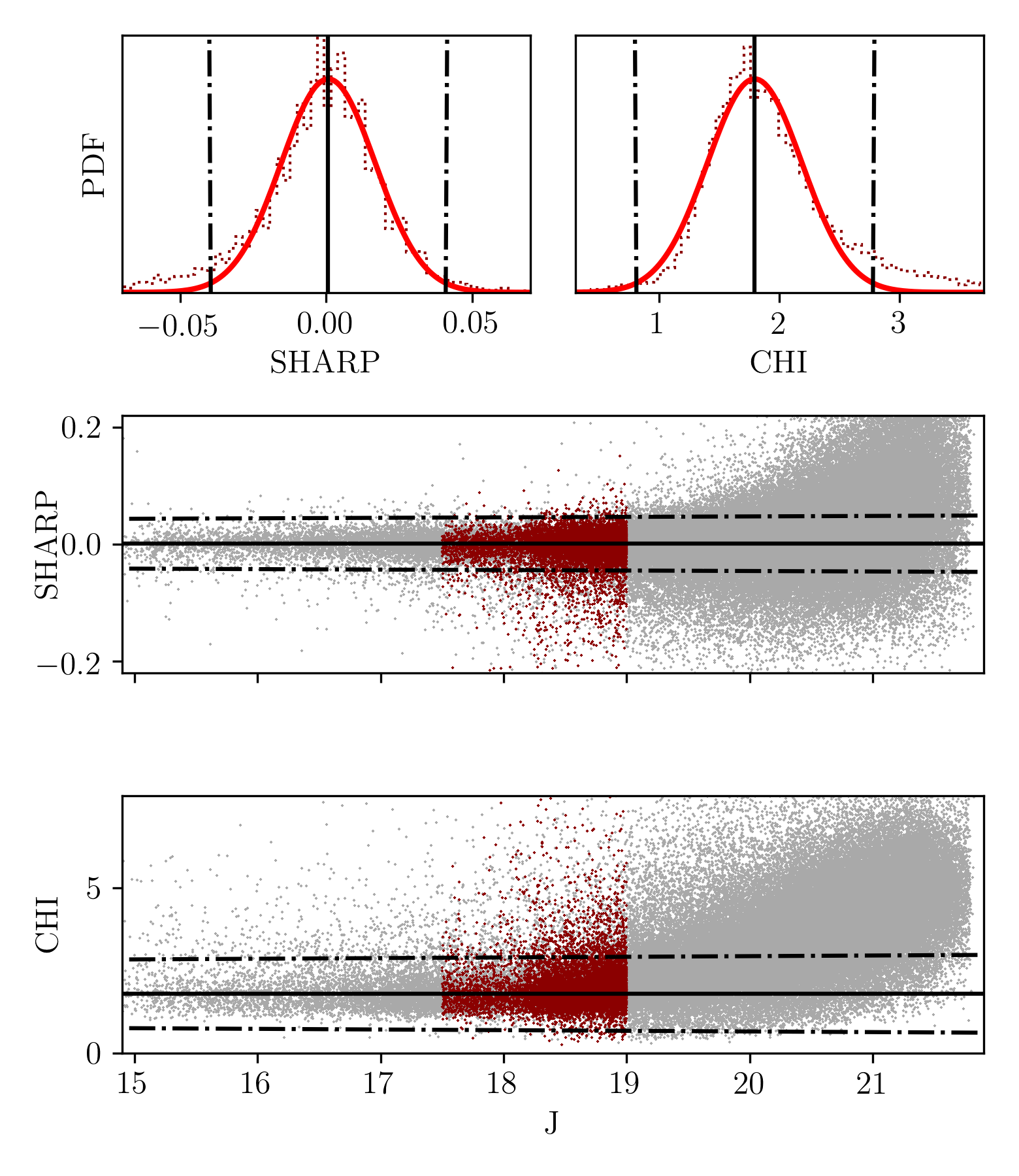}
\caption{The top panels show the distributions of SHARP and CHI values for the selected red giant stars, the vertical continuous lines indicate the mean of the distribution and the other two vertical lines are placed at $\pm 2.5\sigma$ from the mean. The centre and bottom panels show SHARP and CHI as a function of $J$ magnitude. The red dots are the red giant stars selected in the colour-magnitude diagram. The black lines (now horizontal) are again located at the mean (continuous line) and at $\pm 2.5\sigma$ from the mean. The grey dots are all the stars from the left panel of Fig. \ref{fig:ngc6822cmd}}
\label{fig:ngc6822sharpchi}
\end{figure}

To define the mean value of SHARP and CHI in NGC 6822 we need to determine a good reference population. We select stars in the red giant and asymptotic giant stage within a magnitude and colour range where there is little contamination from background and/or foreground sources. The region selected is shown in the left panel of Fig. \ref{fig:ngc6822clean}. These stars form a well-defined sequence in the CMD dominated by stars that are, most likely, members of NGC~6822. This makes them a good reference population to define a typical value of SHARP and CHI for stars from the galaxy. The distributions of SHARP and CHI values for the selected stars are shown in the top panels of Fig. \ref{fig:ngc6822sharpchi}. We set the SHARP and CHI cutoff at the level at which these parameters reach values of 2.5 times the standard deviation from the mean of the distribution. These values are shown graphically in all panels of Fig. \ref{fig:ngc6822sharpchi}. Given the mean and dispersion values of the parameters for the red giants, we define the SHARP and CHI limits for a star to be considered a well-defined point-like source:
\begin{equation}
\begin{split}
&\mathrm{\overline{SHARP}_{RG}} - 2.5\sigma_{RG} < \mathrm{SHARP} < \mathrm{\overline{SHARP}_{RG}} + 2.5\sigma_{RG} \\
&\mathrm{CHI} < \mathrm{\overline{CHI}_{RG}} + 2.5\sigma_{RG}.
\label{eq:sharpchi}
\end{split}
\end{equation}
$\sigma$ represents the standard deviation for SHARP and CHI respectively. As we can see in the lower panel of Fig. \ref{fig:ngc6822sharpchi}, nearly all the stars are above the $2.5\sigma$ limit below the median line, therefore it is not necessary to set a lower limit for CHI.

The cleaned CMD with the stars that fall within the limits defined in Eq. \ref{eq:sharpchi}, is shown in the right panel of Fig. \ref{fig:ngc6822clean}. The SHARP and CHI cutoff helped remove background stars as well as blended stars.


\subsection{IC 1613}
\label{subsec:ic1613phot}

\begin{figure}
\centering
\includegraphics[width=\columnwidth]{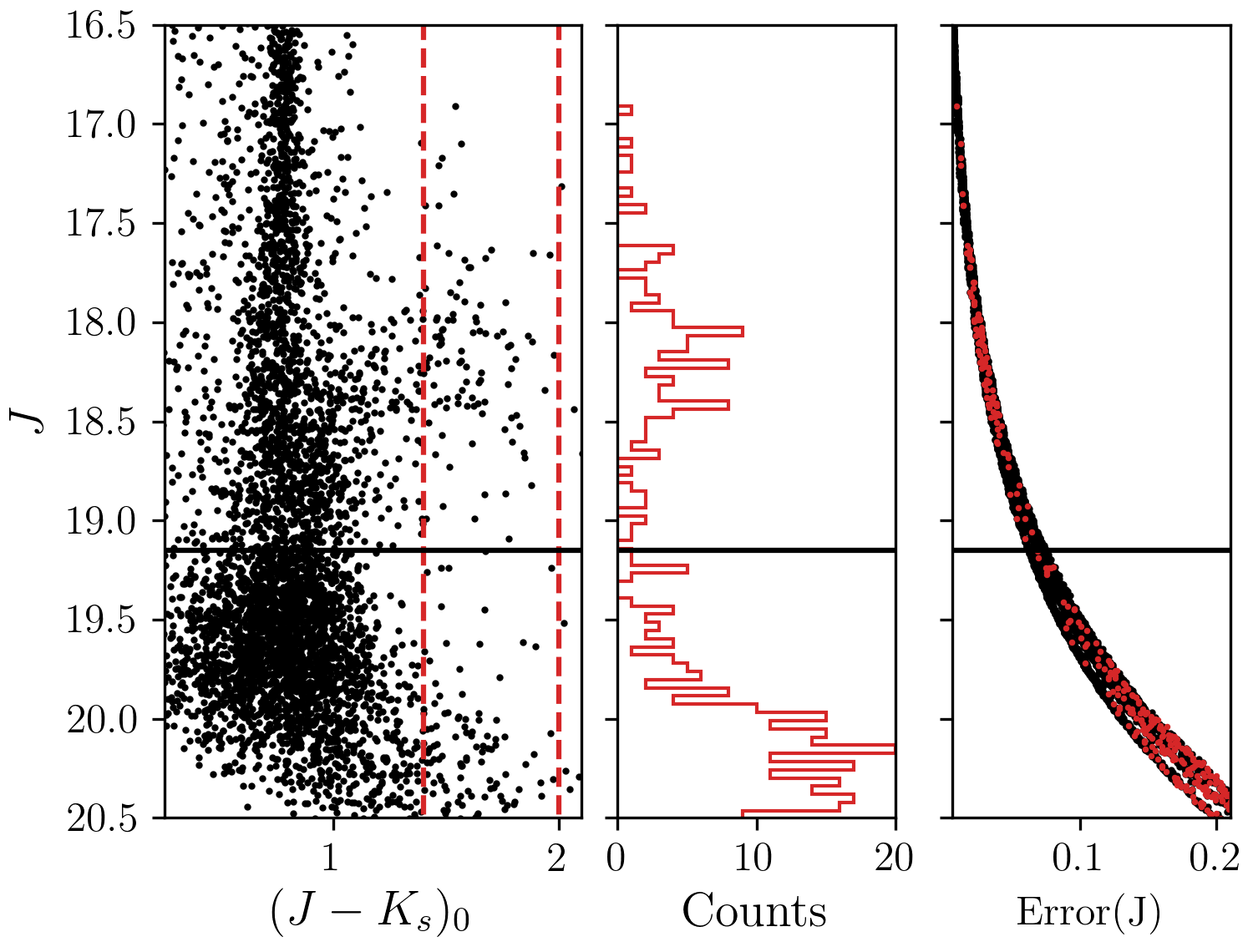}
\caption{\textit{Left:} ($(J-K_{s})_0,J_0$) colour-magnitude diagram for IC 1613. The dotted vertical lines indicate the colour limits for the carbon star region. \textit{Centre:} Luminosity function for all the stars in the CS colour range. \textit{Right:} $J$ magnitude versus the $J$ photometric error. The continuous horizontal line in all three panels marks the faint magnitude cutoff for the CS luminosity function.}
\label{fig:ic1613cmdCS}
\end{figure}

The catalogue for IC 1613 comes from \cite{2015A&A...573A..84S} and is publicly available through VizieR\footnote{\url{http://cdsarc.unistra.fr/viz-bin/cat/J/A+A/573/A84}}. The data are from observations with the Wide Field Camera (WFCAM) on UKIRT in Hawaii. The photometric data provided in this catalogue are calibrated to 2MASS, therefore we only correct for the average line-of-sight extinction $E(B-V)=0.021$ \citep{2011ApJ...737..103S}. For each filter, along with the magnitude for each detected source, the catalogue also provides a numerical flag to categorize the source. There are a total of seven categories of which we only keep those sources designated as "stellar" or "probably stellar" in both $J$ and $K_s$ filters.  The final reddening-corrected CMD for IC 1613 is presented in the left panel of Fig. \ref{fig:ic1613cmdCS}.

\section{Carbon Stars}
\label{sec:CS}

\subsection{Selection}
\label{subsec:CSsel}

In Paper I, by studying the CS in the LMC and SMC, we developed a method to identify these stars based solely in their $(J-K_{s})_{0}$ colour. We catalogue stars as CS if their $(J-K_{s})_{0}$ colour falls in the range:
\begin{equation}
    1.4 < (J-K_{s})_{0} < 2.
\end{equation}
The blue limit assures that the CS are well separated from the oxygen-rich AGB stars. The red limit is to avoid contamination from extreme CS. Extreme (or obscured) CS are CS with larger $(J-K_s)$ colours due to dusty circumstellar envelopes \citep{1996MNRAS.279...32Z, 2000ApJ...542..804N, 2006A&A...447..971V, 2011AJ....142..103B}. Contrary to the relatively constant brightness in the J-band of the CS used in this paper, the J magnitude of extreme CS drops as they get redder.
\removed{These values} \inserted{The values for the colour limits found in Paper I} are in agreement with \cite{2001A&A...377..945C} and \cite{2001ApJ...548..712W}.

When we encounter considerable contamination of non-member stars in the CS colour range, a slight magnitude cut at the faint end may be necessary. It is important to mention that the robustness of our method ensures that a slight magnitude cut does not disrupt the estimation of the distance (see Sec. 2.3 of Paper I and Sections \ref{subsec:lumfunc} and \ref{sec:discussion} of this paper for details).

\subsubsection{Magellanic Clouds}
\begin{figure}
\centering
  \begin{minipage}[b]{0.98\columnwidth}
    \includegraphics[width=\columnwidth]{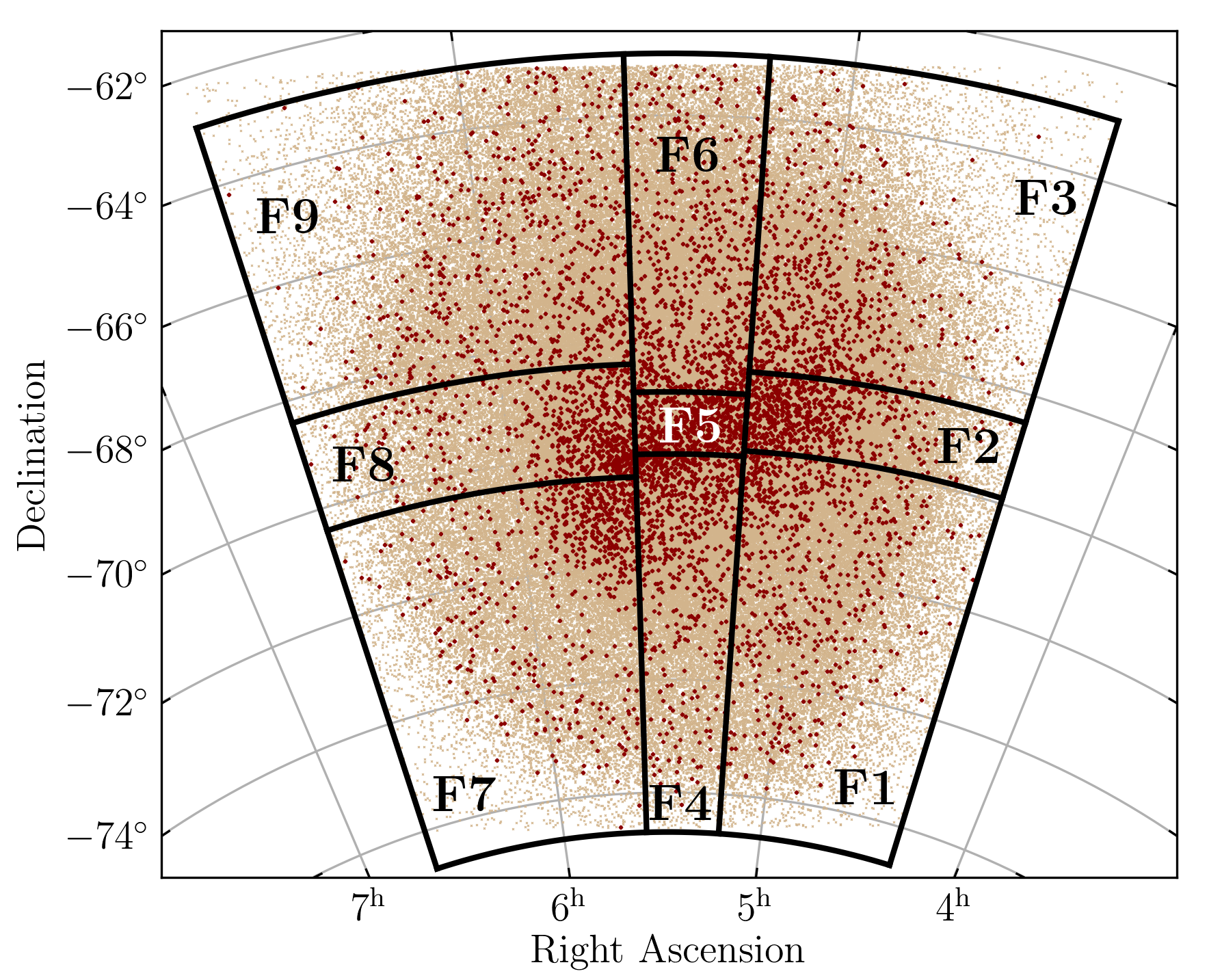}
  \end{minipage}
  \begin{minipage}[b]{1.\columnwidth}
    \includegraphics[width=\columnwidth]{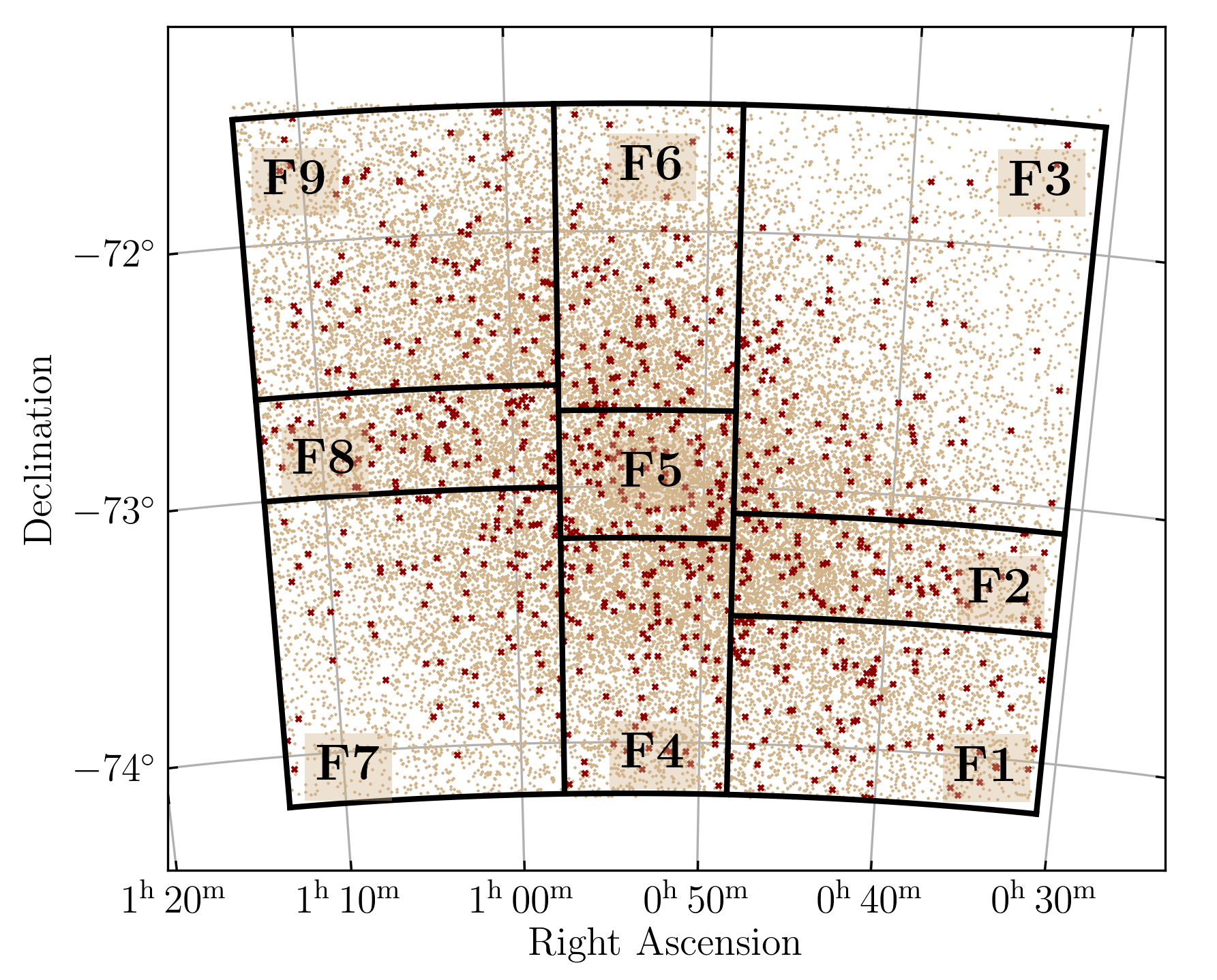}
  \end{minipage}
\caption{\inserted{Right Ascension and declination map for the LMC ({\it top} panel) and the SMC ({\it bottom} panel). The positions of the carbon stars in both galaxies are indicated by the red markers. The black lines delimit the nine fields described in Sec. \ref{sec:3dstruct}, each field contains an equal number of CS (93-94 for the SMC and 667-668 for the LMC).}}
\label{fig:lmc_smc_radec}
\end{figure}

For the LMC and SMC it was possible to obtain a clean CMD by removing the foreground contamination (as detailed in Paper I). In addition, background galaxies are much fainter than the magnitude range observed by 2MASS. Therefore, for both Magellanic Clouds, it is not necessary to define a faint magnitude limit within the CS colour range. Fig. \ref{fig:CSselMCs} shows the CMD and CS selection region for the Magellanic Clouds \inserted{ where we find 6,009 CS for the LMC and 843 CS for the SMC. The distribution on the sky of the CS within the LMC and SMC are shown in the top and bottom panels of Fig. \ref{fig:lmc_smc_radec} respectively.}

\subsubsection{NGC6822}
\begin{figure}
\centering
\includegraphics[width=1\columnwidth]{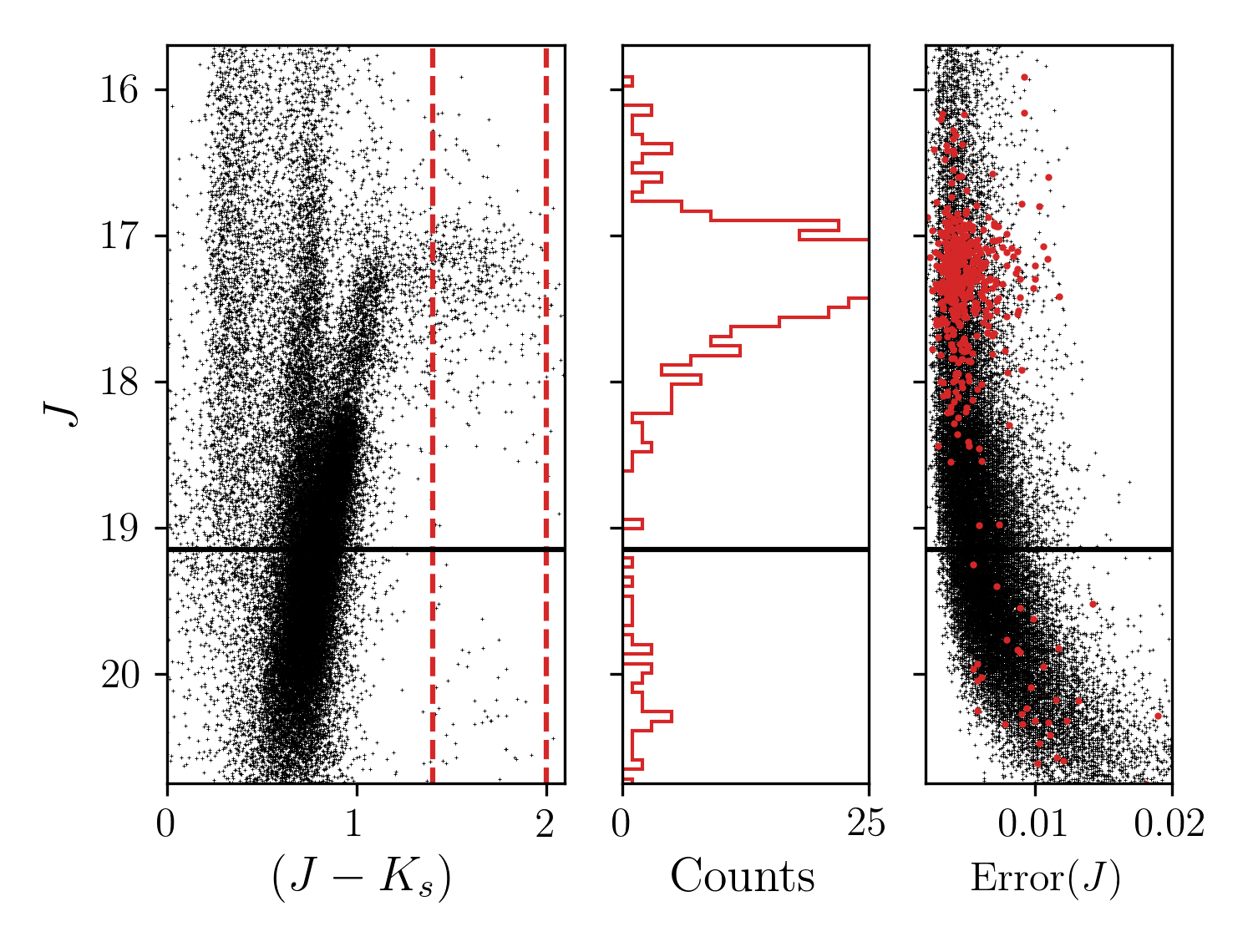}
\caption{Same as Fig. \ref{fig:ic1613cmdCS} but for NGC 6822.}
\label{fig:ngc6822CSmagCut}
\end{figure}

For NGC 6822 we were also able to remove most of background contamination in the CMD. Nevertheless, \inserted{there is a significant number of faint sources that results in a small bump in the luminosity function below magnitude $\sim20$} \removed{when we look at the luminosity function of all the stars in the CS colour range we can still see a small feature at the faint end of the distribution} (see centre panel of Fig. \ref{fig:ngc6822CSmagCut}). These faint stars are not CS but most likely remaining background contamination. As we will see in Sec.~\ref{sec:discussion}, including or removing the stars responsible for this bump does not significantly affect the determination of the distance to NGC 6822. However \removed{they do affect the luminosity function fitting procedure} \inserted{the excess of faint stars in the CS region has an impact on the model fitting procedure, especially for the parameters that shape the faint tail of the luminosity function} (see sub-section \ref{subsec:lumfunc}).

\begin{figure}
\centering
    \includegraphics[width=0.8\columnwidth]{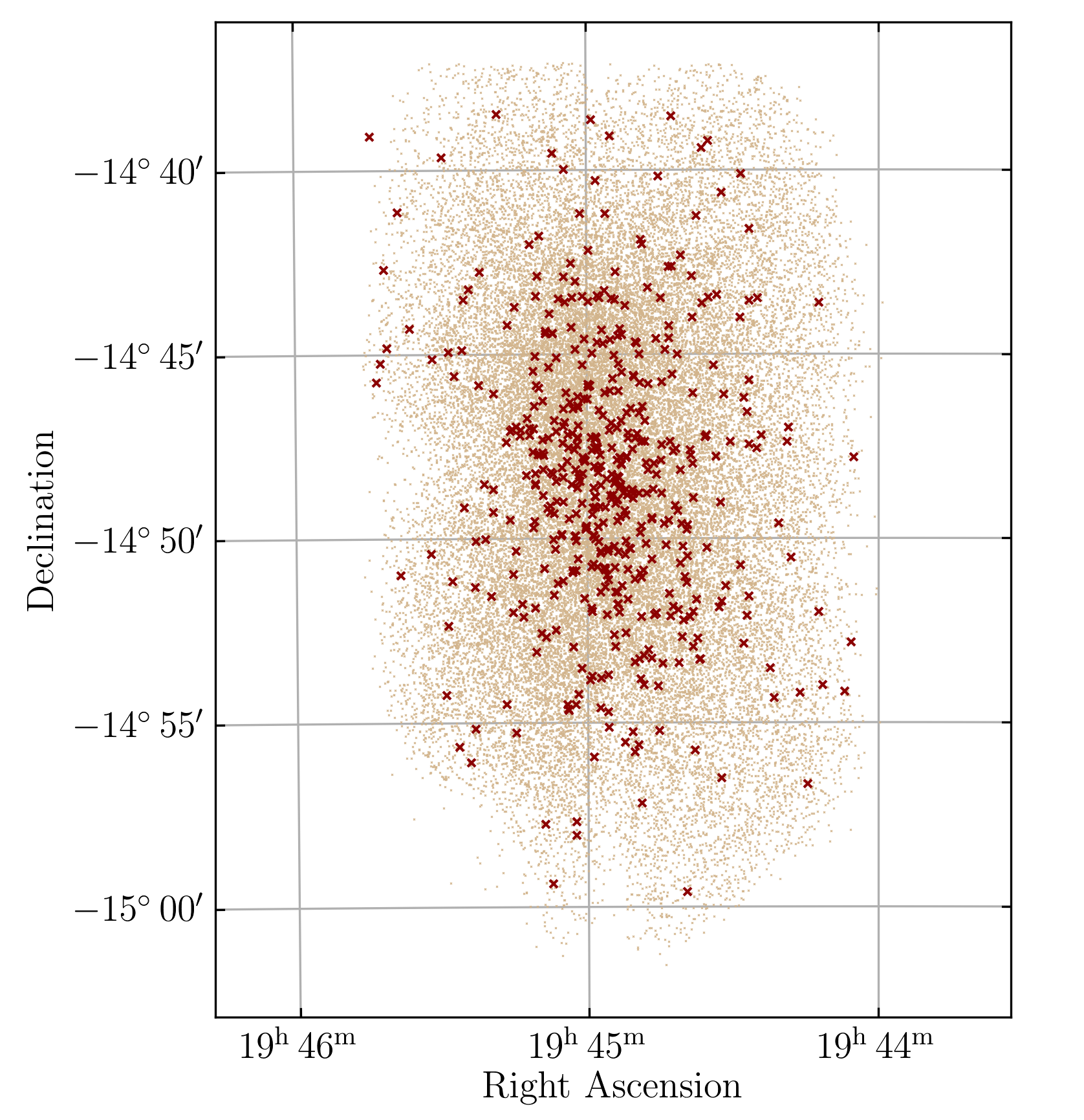}
\caption{\inserted{Right Ascension and declination map for NGC 6822. The positions of the carbon stars in the galaxy are indicated by the red markers.}}
\label{fig:ngc_radec}
\end{figure}

Looking at the histogram in the central panel of Fig. \ref{fig:ngc6822CSmagCut} we can see that there is a relatively flat distribution of stars between the two main features in the CS colour range. We take a point roughly at the centre of the flat region to define the faint magnitude limit for the CS region in NGC 6822 at $J$=19.15 mag. \inserted{We find a total of 484 stars within the limits of the CS region. The location of these stars in the galaxy is shown in Fig. \ref{fig:ngc_radec}.}

\subsubsection{IC 1613}
\begin{figure}
\centering
    \includegraphics[width=1\columnwidth]{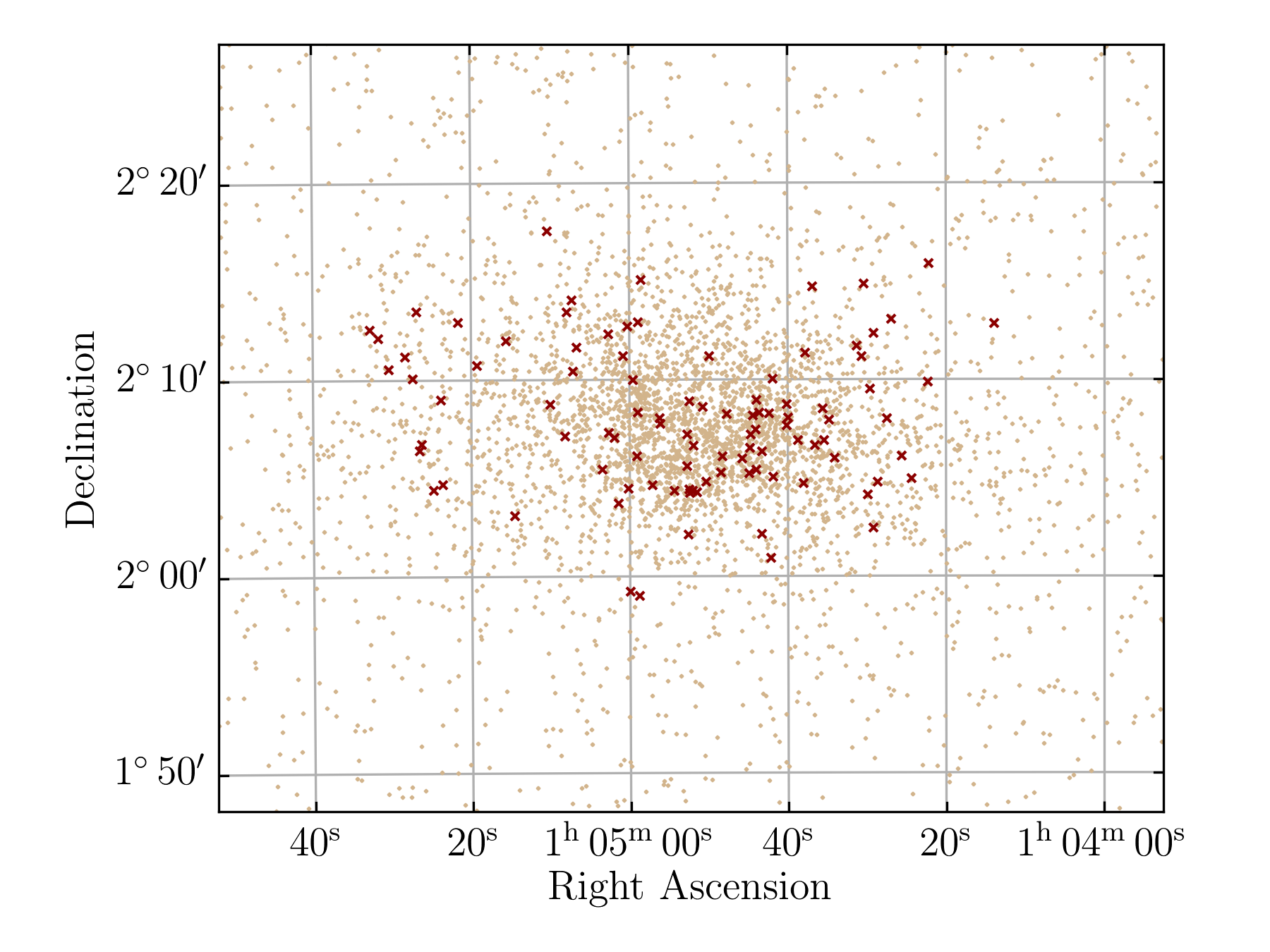}
\caption{\inserted{Same as Fig. \ref{fig:ngc_radec} but for IC 1613. }}
\label{fig:ic_radec}
\end{figure}

In the centre panel of Fig. \ref{fig:ic1613cmdCS},  we note that the luminosity function for the stars in the CS colour range is highly contaminated at the faint end. Therefore, for IC 1613, we take the same approach as for NGC 6822 to define a faint magnitude limit for the CS region. As shown in Fig. \ref{fig:ic1613cmdCS} the faint limit is set at $J$=19.20 mag. \inserted{Within these limits we find a total of 99 CS whose positions within the galaxy are indicated in Fig. \ref{fig:ic_radec}.}

\subsection{ Luminosity function}
\label{subsec:lumfunc}

\begin{figure*}
\centering
\includegraphics[width=\textwidth]{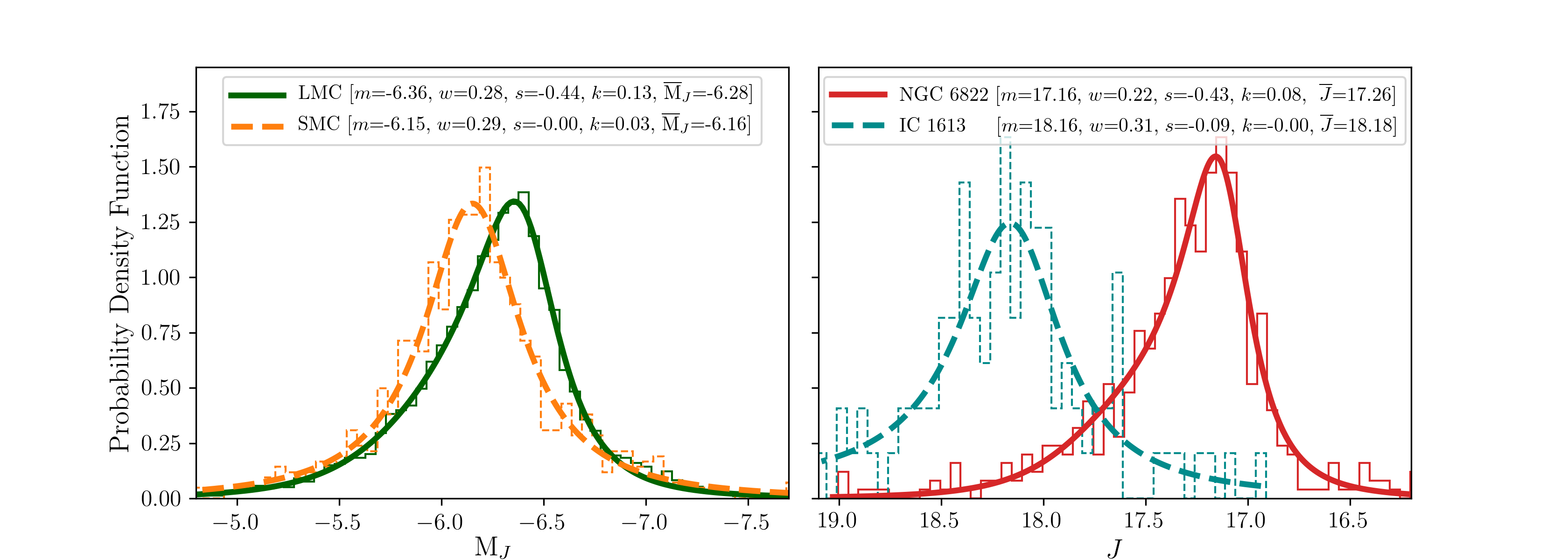}
\caption{Best-fit modified Lorentzian function for the carbon star luminosity functions for the LMC (solid green line) and SMC (dashed orange line) in the \textit{left} panel, and for NGC 6822 (solid red line)  and IC 1613(dashed dark-cyan line) in the \textit{right} panel. The inset specifies the values for the best-fit parameters for Eq. \ref{eq:lfmodel} and ($\overline{J}$) for each galaxy. The histograms behind each model curve depict the actual data for each galaxy.}
\label{fig:lumfuncs}
\end{figure*}

In Paper I we examined the CS luminosity function for the Magellanic Clouds and the Milky Way. For each of these galaxies we estimated the median absolute $J$ magnitude ($\mathrm{\overline{M}}_{J}$) and its dispersion ($\sigma_{\mathrm{\overline{M}}_{J}}$) using maximum likelihood statistics (see Sec. 2.3 of Paper I for details). In this initial study we assumed that (for any galaxy) the $J$ magnitudes of the CS, in the given $1.4<(J-K_s)_0)<2.0$ colour range, are normally distributed about ($\mathrm{\overline{M}}_{J}$). \inserted{However, a normal distribution does not describe the shape of the CS luminosity functions. As can be seen in Fig. \ref{fig:lumfuncs}, in all four galaxies the LFs exhibit extended tails and are slightly asymmetric.}

In the present study we model the luminosity functions of the CS using a modified Lorentzian distribution function with two extra parameters to allow for asymmetry (skewness) and to vary the weight of the tails (kurtosis). The CS luminosity function model is defined as:
\begin{equation}
    f(J;m,w,s,k) = \frac{a}{1 + \left( \frac{J-m}{w} \right)^2 + s\left( \frac{J-m}{w} \right)^3 + k\left( \frac{J-m}{w} \right)^4},
\label{eq:lfmodel}
\end{equation}
where:
\begin{itemize}[leftmargin=*,label={}]
    \item $m$ = mode of the distribution,
    \item $w$ = scale parameter (specifies the width),
    \item $s$ = skewness parameter,
    \item $k$ = kurtosis parameter,
    \item $a$ = amplitude (height of the peak).
\end{itemize}
We allow the skewness and kurtosis parameters to take positive or negative values, so the functional form may have minima far from the peak (in particular when $9 s^2>32 k$); consequently, we apply Eq.~\ref{eq:lfmodel} only over the modest range of magnitudes that carbon stars exhibit.  We fit the model from Eq. \ref{eq:lfmodel} to the LMC, SMC, NGC 6822 and IC 1613. The best-fit curves are over-plotted on the data in Fig. \ref{fig:lumfuncs}. We repeat the best-fitting procedure over ten-thousand bootstrap realizations for each galaxy. The values and errors for the luminosity function parameters from the bootstrapping procedure are summarized in Tab. \ref{tab:best-fit_lmcsmc} for the LMC and SMC and Tab. \ref{tab:best-fit_ngcic} for NGC 6822 and IC 1613. 

\inserted{From the bootstrapping results we find that the median $J$ magnitude does not correlate with any of the parameters from the Lorentzian function. Among the model parameters, the width correlates with the kurtosis and has an anti-correlation with the amplitude. This is expected given that a sample of CS with less dispersion in the magnitude should have a more pronounced LF peak. At the same time, given that the magnitude range remains constant, a wider distribution should also show more extended tails. We also find an anti-correlation between the skewness and the mode. The relation between these parameters is logical given the constant behaviour of the median; in other words, if the mode moves away from the median, the skewness increases negatively or positively depending on the direction of the difference between the mode and the median.}

\inserted{Using a Lorentzian for the fitting has the advantage that the mode is less sensitive to the magnitude selection compared to models that fit for the mean. An abrupt magnitude cut that prevents the luminosity function from smoothly declining will perturb the fitting of the tails of the distribution, affecting the kurtosis parameters. Even if the kurtosis parameter is affected by the location of the peak, the mode and skewness are not affected. As we will see in Sec. \ref{sec:distance},  the mode and skewness are the most important parameters in the distance determination method. These only correlate with each other and are not significantly affected by the magnitude selection.   }

The new model allows us to see that, not only is ($\mathrm{\overline{M}}_{J}$) different for the LMC and SMC, but, as can be seen in Fig. \ref{fig:lumfuncs} the shape of the luminosity function also varies from galaxy to galaxy. For instance, for the LMC  and NGC 6822 the luminosity functions are skewed towards brighter magnitudes while for the SMC and IC 1613 the functions are more symmetric.

\begin{table}
\caption{Large and Small Magellanic Clouds median absolute J magnitudes and best-fitting parameters for the CS luminosity function model specified in Eq. \ref{eq:lfmodel}. Values and errors are obtained from 10000 bootstrap realizations.}
\centering
\begin{tabular}{l r r}
\hline
& \multicolumn{1}{c}{LMC} & \multicolumn{1}{c}{SMC} \\
\hline
\hline
\vspace{0.16cm}
Median  & -6.283$^{+0.004}_{-0.005}$ & -6.160$^{+0.013}_{-0.016}$   \\
\vspace{0.16cm}
Mode & -6.36$^{+0.01}_{-0.01}$ & -6.15$^{+0.02}_{-0.02}$ \\
\vspace{0.16cm}
Width &  0.28$^{+0.01}_{-0.01}$ & 0.29$^{+0.03}_{-0.03}$  \\
\vspace{0.16cm}
Skew & -0.43$^{+0.06}_{-0.06}$ & -0.01$^{+0.07}_{-0.08}$ \\
\vspace{0.16cm}
Kurtosis &  0.13$^{+0.03}_{-0.02}$ & 0.03$^{+0.04}_{-0.02}$   \\
Amplitude &  1.34$^{+0.03}_{-0.03}$ & 1.34$^{+0.08}_{-0.07}$   \\
\hline
\end{tabular}
\label{tab:best-fit_lmcsmc}
\end{table}
%

\begin{table}
 \caption{Same as Tab. \ref{tab:best-fit_lmcsmc} for NGC 6822 and IC 1613.}
 \centering
 \begin{tabular}{l r r}
 \hline
 & \multicolumn{1}{c}{NGC 6822} & \multicolumn{1}{c}{IC 1613} \\
 \hline
 \hline
\vspace{0.16cm}
Median & 17.26$^{+0.03}_{-0.02}$ & 18.18$^{+0.02}_{-0.07}$   \\
\vspace{0.16cm}
Mode &  17.16$^{+0.03}_{-0.03}$  &  18.16$^{+0.13}_{-0.09}$ \\
\vspace{0.16cm}
Width &   0.23$^{+0.04}_{-0.04}$  &   0.25$^{+0.10}_{-0.11}$ \\
\vspace{0.16cm}
Skew &  -0.49$^{+0.15}_{-0.30}$  &  -0.04$^{+0.42}_{-0.42}$ \\
\vspace{0.16cm}
Kurtosis &   0.11$^{+0.16}_{-0.05}$  &  -0.03$^{+0.16}_{-0.07}$ \\
Amplitude & 1.56$^{+0.14}_{-0.11}$  &   1.36$^{+0.37}_{-0.21}$ \\
 \hline
 \end{tabular}
 \label{tab:best-fit_ngcic}
\end{table}

\subsection{\inserted{Median $J_0$ magnitude}}
\label{sec:medianJ}

Paper I details how we estimate the median $J_0$ magnitude of the CS using a robust estimator that is less sensitive to outliers. These outliers are likely massive oxygen rich stars (at the bright end) and background or foreground contamination (at the faint end). The analysis carried out in Sec. \ref{sec:discussion} shows that the effect on the median $J_0$ magnitude of a magnitude selection (i.e. the inclusion or exclusion of outliers) is minimal. Nevertheless, there are other sources of uncertainty that can affect the median $J_0$ magnitude.

\inserted{In order for the median $J_0$ magnitude to be considered a suitable standard candle, it is essential that we can account for the small differences in its value from galaxy to galaxy. Metallicity and age  may have an effect on the magnitude of the stars. In order to asses the effect of these properties, we compare the CS in the SMC with thermally pulsing AGB (TP-AGB) theoretical isochrones with different initial metallicites and at different ages.}

\begin{figure}
\centering
    \includegraphics[width=1\columnwidth]{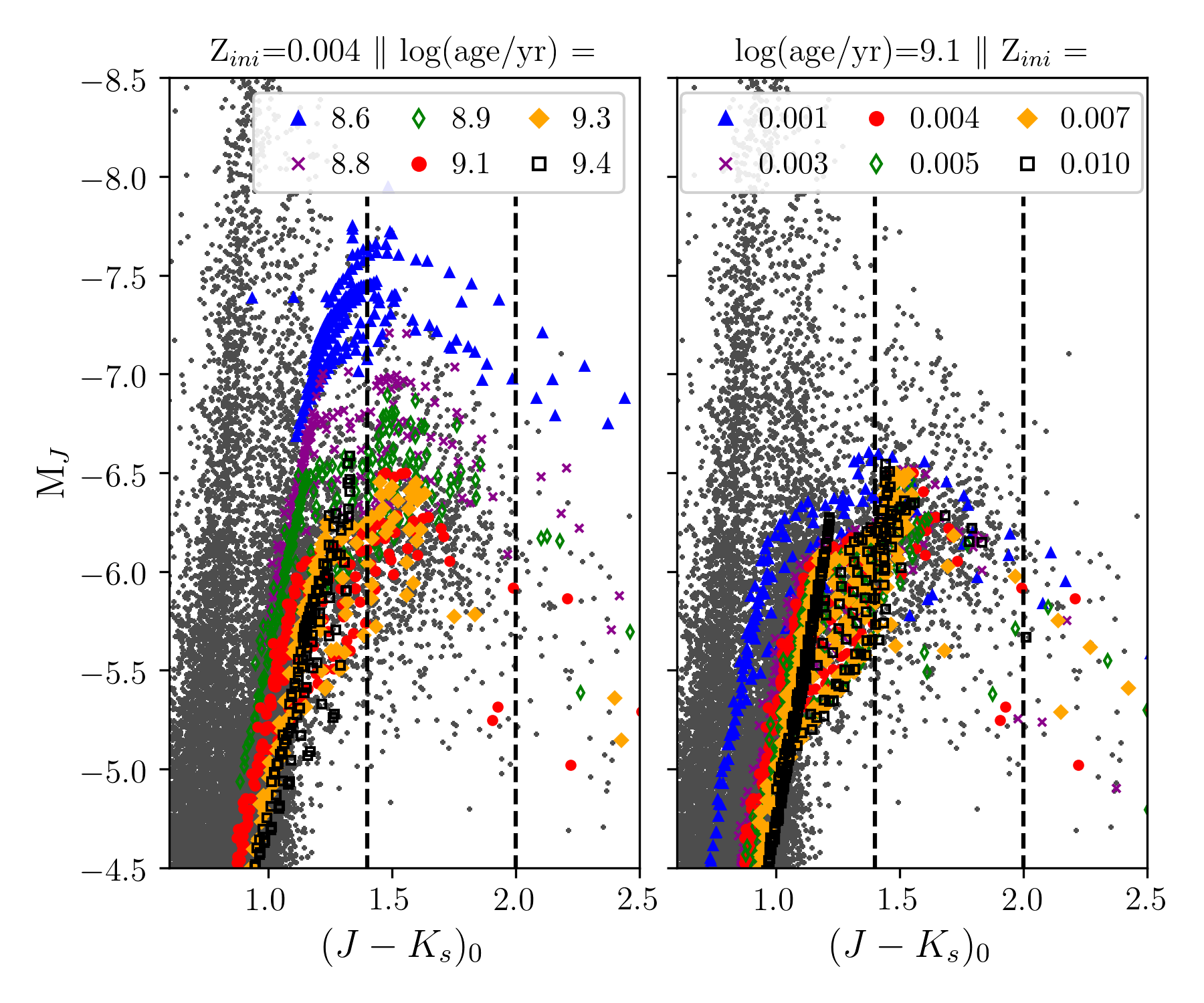}
\caption{\inserted{\texttt{COLIBRI} isochrones for thermally pulsing asymptotic giant branch stars (TP-AGB) plotted over the ($(J-K_{s})_0,$M$_J$) colour-magnitude diagram of the SMC. {\it Left:} The different marker types$/$colours represent isochrones of different ages as specified in the inset. All the isochrones are generated using an initial metallicity value of $Z_{ini}=0.004$. {\it Right:} The different marker types$/$colours represent isochrones of log(age$/$yr)=9.1 generated with different initial metallicities as specified in the inset.} For all isochrones (except for log(age$/$yr)=8.9) there are no oxygen rich AGB stars in the color range demarcated by the dashed black vertical lines. All the model stars with C$/$O$>$1 have at least $(J-K_{s})_0$>1.2. }
\label{fig:smc_iso}
\end{figure}

\inserted{ \cite{2019MNRAS.485.5666P, 2020MNRAS.498.3283P} used detailed star-formation histories for the SMC (from \citealt{2018MNRAS.478.5017R}) and the LMC in order to characterize the TP-AGB population of these galaxies. Their work allowed them to improve the constrains on the physical parameters that dominate the evolution of the stars at this stage of evolution. These new constrains on the TP-AGB phase were incorporated into the \texttt{PARSEC-COLIBRI} stellar isochrones\footnote{Publicly available at \url{http://stev.oapd.inaf.it/cmd}} \citep{2012MNRAS.427..127B, 2017ApJ...835...77M}. We generated \texttt{PARSEC-COLIBRI} theoretical isochrones with \texttt{PARSEC 1.2S} and \texttt{COLIBRI S\_35} options, following the best-fit results for the SMC from \citet{2019MNRAS.485.5666P}. Fig. \ref{fig:smc_iso} shows the ($(J-K_{s})_0,$M$_J$) colour-magnitude diagram of the SMC with isochrones of different ages (left panel) and different metallicities (right panel). To simulate the bulk of the CS population in the SMC, isochrones of different ages are generated with $Z_{ini}=0.004$  and isochrones of different metallicities are all set to log(age$/$yr) = 9.1. The range of age and metallicity are selected in order to populate the area of the CMD where we see SMC CS. In the plot, we include all the stars labeled as TP-AGB in the isochrone models. Further inspection of the isochrone data shows that the area within the defined colour limits for CS is populated predominantly by stars with C$/$O$>$1. For all models, except for the youngest isochrone (log(age$/$yr) = 8.6), we find that oxygen rich stars fall on the blue side of our colour range and all stars with C$/$O ratios greater than one have $(J-K_{s})_0$ colour greater than 1.2. } 

\inserted{As reflected in Fig. \ref{fig:smc_iso}, the theoretical isochrones are consistent with the locations of the real stars. The bulk the CS from the SMC in the colour range specified in Sec. \ref{subsec:CSsel} fall in the range of the isochrones with log(age$/$yr) between 8.9 and 9.3. We can see that older stars are bluer than the blue limit of our CS selection colour range. The younger isochrones cross through the CS area but, as seen in the figure, there are not many SMC CS populating that region of the CMD. The effects of metallicity on the magnitude of CS, as seen in the right panel of Fig. \ref{fig:smc_iso}, are less pronounced than the change in magnitude with age. From \cite{2019MNRAS.485.5666P} we know the CS population of the SMC is largely populated by stars with $Z_{ini}=0.004$; nevertheless the presence of CS with different metallicities does not produce a large dispersion in magnitude within the CS colour range.}

\inserted{Another possible source of uncertainty in the median $J$ magnitude is stellar variability. When we measure the distribution of magnitudes of the CS we are including stars that might be inherently different and$/$or stars that could potentially be identical but are at different phases in their variation cycle. With single epoch observations it is impossible to determine the precise origin of the magnitude dispersion. However, unlike distance determination methods that depend on star by star measurement over a variation cycle (e.g. RR Lyrae and Cepheid variables), our method relies on the ensemble of all the CS within the given colour range.  This ensemble is an average over the different types of stars but also the same star at different parts of its cycle. Therefore, we expect the effects of variability on the median magnitude to average out. The fact that only a single epoch of observations are needed is one of the advantages of this method over the use of variable stars.}


\subsection{Metallicity and star formation history effects on the luminosity function}
\label{sec:ZandSFH}

It is clear from Fig. \ref{fig:lumfuncs} that the luminosity function of the CS is different for different galaxies. From Table \ref{tab:best-fit_lmcsmc}, the main differences between the CS luminosity function of the LMC and SMC are the mode and the skew parameter. The LMC distribution of CS is skewed towards brighter magnitudes while for the SMC the distribution is more symmetric and it peaks at a lower magnitude. \inserted{Various studies have linked the differences in the parameters of the CS luminosity functions to intrinsic global properties of the host galaxy such as metallicity and star formation history (SFH).}


\subsubsection{Metallicity}
\label{sec:met}

There are a vast number of studies on the metallicity of the LMC and SMC. Most studies agree the LMC has a metallicity gradient and on average a higher metallicity than the SMC (see \citealt{2016MNRAS.455.1855C, 2018MNRAS.475.4279C} and references therein). Recently, \cite{2016MNRAS.455.1855C, 2018MNRAS.475.4279C}, created high-resolution metallicity maps of the LMC and SMC using two large photometric surveys: the Magellanic Cloud Photometry Survey (MCPS) and the Optical Gravitational Lensing Experiment (OGLE III), together with spectroscopic data of red giant stars. They found shallow radial metallicity gradients for both the LMC and SMC with average metallicities of $\left<[\mathrm{Fe/H}]\right>\sim -0.38$ and $\left<[\mathrm{Fe/H}]\right>\sim -0.95$ for the LMC and SMC respectively.

\inserted{AGB stars have also been used to estimate metallicity.} Because the transition from oxygen-rich to carbon-rich AGB stars depends on metallicity, the ratio between CS stars and spectral type M AGB stars (C/M ratio) can be used as a tracer of metallicity in local group galaxies: e.g.  \cite{1983ApJ...264..114R, 1985ApJ...298L..13R, 1987ApJ...323...79P, 1995AJ....109.2480B, 2003A&A...402..133C, 2008A&A...487..131C, 2009A&A...506.1137C, 2010MNRAS.408L..76F, 2015ApJ...810...60H}. Among the numerous publications, \cite{2009A&A...506.1137C} 
found variations in the C/M ratio of the LMC and SMC (and M33). The authors updated the $[\mathrm{Fe/H}]$ vs. log(C/M) calibration from \cite{2005A&A...434..657B} and used it to convert observed C/M ratios into metallicity values. Their results showed that the metallicity in the LMC decreases radially with a peak in the metallicity distribution at $[\mathrm{Fe/H}]\sim-1.15$ while they found no clear trend for the SMC and a constant value of $[\mathrm{Fe/H}]\sim-1.25$.

For our two test galaxies, NGC 6822 and IC 1613, we also find a wide range of values for [Fe/H]. For NGC 6822 \cite{2001ApJ...547..765V} found $\left<[\mathrm{Fe/H}]\right>=-0.49\pm0.22$ using the spectra of A-type supergiants. \cite{2016MNRAS.456.4315S} used Ca II triplet spectra of red giant stars and found $\left<[\mathrm{Fe/H}]\right>=-0.84\pm0.04$. Recently, \cite{2020ApJ...892...91H} found a global $[\mathrm{Fe/H}]=-1.286\pm0.0956$ using AGB stars. See Table 3 of \cite{2016MNRAS.456.4315S} for a summary of other [Fe/H] values for NGC 6822. For IC 1613, \cite{2003ApJ...596..253S} used $HST$ photometry to derive a broad metallicity range $-1.3<[\mathrm{Fe/H}]<-0.7$ and evidence of constant star formation. Using the spectra of red giants \cite{2013ApJ...779..102K} found $[\mathrm{Fe/H}]=-1.19\pm0.01$. Using the C/M ratio \cite{2015A&A...573A..84S} and \cite{2015A&A...578A..51C}
obtained [Fe/H$]=-1.26\pm0.07$ and [Fe/H$]=-1.23\pm0.06$ respectively. 

\inserted{Theoretical studies have shown a connection between the host galaxy's metallicity and the shape of its CS luminosity function.} \cite{1999A&A...344..123M} reproduced the observed luminosity functions of the CS in the Magellanic Cloud using synthetic AGB star models. They concluded that the difference in the location of the peak of the luminosity function is due to the difference in metallicity between the two galaxies, specifically how metallicity affects the third dredge-up (3DU) efficiency parameter. The authors found that the difference in the peak is reproduced by the SMC having a higher efficiency. This is in agreement with other theoretical studies that indicate that dredge-up efficiency increases as metallicity decreases (e.g. \citealt{2002PASA...19..515K, 2015ApJS..219...40C, 2017ApJ...835...77M, 2019ApJ...879..109B}).

\inserted{\cite{2019MNRAS.485.5666P, 2020MNRAS.498.3283P} find that the current population of CS in the LMC is best represented as having a metallicity of $Z_{ini}=0.008$ while for the SMC $Z_{ini}=0.004$, in agreement with the LMC having, on average, a higher observed metallicity. Their best models also exhibit a lower efficiency in the 3DU parameter for the LMC. They explain that one of the consequences of a lower efficiency is that the C$/$O $>$ 1 transition takes place at brighter magnitudes. }

\inserted{Through its effect on the 3DU efficiency parameter,} metallicity plays a role in defining the shape of the CS luminosity function \inserted{with higher metallicities shifting the luminosity function peak to brighter magnitudes.} But this is a parameter for which defining a global value for a galaxy is not straightforward. As we can see from the various  values quoted above, the average metallicity depends on the method and stellar population used to measure it. \inserted{As a result, the global metallicity of a galaxy may not necessarily be representative of the metallicity of the bulk of the CS population.}

\subsubsection{Star formation history}

\inserted{SFH may also play an important role in the shape of the CS luminosity function.} \cite{1999A&A...344..123M} claim that the drop at bright magnitudes in the CS luminosity function of the LMC can be attributed to a recent decrease in the star formation rate (SFR). \inserted{Here we summarize the most important global aspects of the SFH for the four galaxies included in this study.}

\inserted{The LMC and SMC have been the target of numerous SFH studies. \cite{2009AJ....138.1243H} and \cite{2014MNRAS.438.1067M} find similar results for the global SFH of the LMC with the main events at slightly different ages. Both studies report an early burst followed by a quiescent period that lasted several billion years. After this inactive period, around $\sim5$ Gyr ago, a second episode of star formation took place characterized by bursts localized in specific regions of the galaxy. The age-metallicity relation (AMR) is found to reflect the SFH. With an early epoch of chemical enrichment, constant metallicity values during the quiescent period, and an steady increase in metallicity after the re-ignition of star fomration. }

\inserted{The SFH of the SMC  is characterized by different aged populations having different spatial distributions \citep{2004AJ....127.1531H, 2013ApJ...775...83C, 2018MNRAS.478.5017R}. Older populations seem to have a smooth spatial distribution, whereas recent star formation is limited to specific areas of the galaxy. Similarly to the LMC, \citet{2004AJ....127.1531H} and \citet{2018MNRAS.478.5017R} find a quiescent period of star formation between $\sim 8$ and $\sim3$ Gyr ago. 
The AMR derived by \citet{2018MNRAS.478.5017R} shows a considerable increase in metallicity between 13 and 1.5 Gyr reaching Z$\sim$0.003 followed a slow increase until 130 Myr ago when it plateaus at present day values ( Z$\sim$0.004).  }

\inserted{\cite{1996AJ....112.1950G, 1996AJ....112.2596G} carried out one of the first attempts to construct an overall SFH for NGC 6822. The authors found a constant or slightly declining SFR over the last few Gyr followed by an enhancement in the star formation activity between 100-200 Myr ago. \cite{2001AJ....122.1807V} found an overall constant or slightly increasing SFR with a possible period of low activity between 3 to 5 Gyr ago. 
\cite{2012ApJ...747..122C} found that over 50\% of the stars in the galaxy were formed in the last $\sim$5 Gyr with an increase in the SFR in the last 50 Myr. The results of \citet{2014A&A...572A..26F} are generally consistent with \citet{2012ApJ...747..122C}, but they find a slow decrease in the SFR in some fields over the past Gyr. Additionally, \citet{2014A&A...572A..26F} are able to derive the AMR for the galaxy finding an increase in metallicity with time throughout the entire galaxy.  }

\inserted{For IC 1613, \citet{2014ApJ...786...44S} derived the mean SFH of the galaxy using one HST$/$ACS field centered at the half-light radius of the galaxy. Their findings indicate a constant SFH over the lifetime of the galaxy with variations of $\sim$30\% or less in the SFR. The continuous star formation in this galaxy is accompanied by a smooth built-up in metallicity over time. }

\subsubsection{Discussion}

 \inserted{\citet{2019MNRAS.485.5666P} used the AMR derived from the SFH to find the initial metallicity distribution for the TP-AGB stars in the SMC and LMC. They successfully modeled the luminosity function of AGB stars in the SMC by calibrating synthetic TP-AGB evolution models. For the LMC, the best-fitting models calibrated for the SMC showed an excess of faint stars, a deficit of stars in the brighter magnitude bins, and an overestimation in the number of CS. \citet{2020MNRAS.498.3283P} resolved the discrepancy between observations and synthetic population in the LMC by adjusting the 3DU parameters in the stellar evolution models. First, the authors increased the value of the core mass for which the efficiency parameter reaches its maximum value. This results in a decrease in the number of faint stars. Second, they decreased the maximum efficiency of the 3DU which, as mentioned previously, shifts the peak of the distribution to brighter magnitudes.}

\inserted{We cannot determine at this time whether the shape of the CS luminosity function is driven mainly by metallicity or SFH. In \citet{2019MNRAS.485.5666P, 2020MNRAS.498.3283P}, SFH is an essential part of their analysis to derive evolutionary models capable of reproducing the observed TP-AGB populations of the SMC and LMC. However, the key parameters that defined the difference between the best-fitting models for the LMC and SMC are the 3DU parameters, which as mentioned in section \ref{sec:met}, are strongly related to metallicity. }

\inserted{As we will discuss in the following sections}, if the distance to a galaxy is measured against a calibrator with a \inserted{significantly} different metallicity \inserted{(or SFH)} we may introduce a systematic error in the distance determination. In Paper I we explored the possibility of using a composite magnitude to account for metallicity effects (see Sec. 5.2 of Paper I). This proved useful to bring the values of ($\mathrm{\overline{M}}_{J}$) for the LMC and SMC closer into agreement but it does not account for the difference in the skew parameter. \removed{In the next section we will show how we determine the proper calibrator, LMC or SMC, depending on the galaxy to which we are measuring the distance} \inserted{In Sec. \ref{sec:distance} we discuss how we determine the most suitable calibrator based on the properties of the CS luminosity function of the target galaxy.}


\section{Distance Determination}
\label{sec:distance}

\subsection{The method}
\label{subsec:method}

\subsubsection{\inserted{Determining the suitable calibrator}}

\inserted{A fundamental step in our method is the determination of the most suitable calibrator for each target galaxy; that is to say, whether a galaxy is "LMC-like" or "SMC-like". Our first attempts were to compare global metallicities and SFHs to find commonalities between the calibrators and the target galaxies but, as mentioned in  \ref{sec:ZandSFH}, these properties do not seem to be representative of the CS populations. Detailed analysis of TP-AGB stellar population, such as the ones carried out by  \cite{2019MNRAS.485.5666P, 2020MNRAS.498.3283P} for the Magellanic Clouds, would be needed in the target galaxies in order to use SFH as the basis for classification. Instead, we believe that it is the observed {\it shape} of the luminosity function that tell us something fundamental about the properties of the galaxies. For example, the skewness could be an indication of the value of the 3DU efficiency parameter. }

\inserted{To ascertain whether it is more appropriate to use the LMC or SMC as the calibrator we classify the target galaxy as "LMC-like" or "SMC-like" based on the shape parameters of the CS luminosity function itself. In section \ref{sec:CS} we mentioned that the main differences between the CS luminosity functions are the magnitude at which the distribution peaks (i.e. the mode of the distribution) and the symmetry of the distribution. There are a few parameters from the Lorentzian model that can be used to asses the differences in the distributions. For example, the variation between the values of the mode and the median changes; the smaller the difference between the mode and the median, the more symmetric the distribution. Another indicator of the shape is the skew parameter; the deviation from zero of the skewness indicates how asymmetric the LF is, while the sign tells whether the LF is inclined towards faint (positive skewness) or bright (negative skewness) magnitudes.} 

\inserted{In principal, we could use the difference between the mean and median, as it is also an indicator of how asymmetric the distribution is. However, the mean is highly sensitive to outliers and variations in the limits defined for the CS region (see Sec. \ref{sec:mVSm}). Using the parameters from the Lorentzian model is preferable, as this method is less sensitive to those variations and is able to identify the mode of the distribution in a robust manner.}

\inserted{The mode can be well defined by the Lorentzian model but by itself does not give enough information. We will therefore employ the skew parameter. The skew parameter is mostly defined by the structure of the LF near the peak of the distribution. This parameter has the advantage that it is not sensitive to variations in the magnitude limits.}


\subsubsection{\inserted{Distance modulus}}

To find the true distance modulus of a given galaxy (let us call it \textit{Gal. X}) we use the median $J$ magnitude ($\mathrm{\overline{J}}$) of the CS luminosity function and compare it to the median of the absolute $J$ magnitude ($\mathrm{\overline{M}}_{J}$) of our calibrator galaxy. \inserted{We define the distance modulus as:}
\removed{In section \ref{sec:ZandSFH} we mentioned that the main effects of metallicity on the luminosity function are the change in the magnitude at which the distribution peaks (i.e. the mode of the distribution) and the shape of the distribution. Without knowing the distance to the galaxy we cannot use the mode by itself to classify a galaxy as LMC or SMC like. We do see that the difference between the values of the mode and the median changes between the LMC and SMC giving an indication of the shape of the distribution (the smaller the difference between the mode and the median, the more symmetric the distribution). Another indicator of the shape is the skew parameter. Focusing on this parameter }
\begin{equation}
    \mu_{0,X} =
\begin{cases}
    \overline{J}_{X} - \mathrm{\overline{M}}_{J_{LMC}},& \text{if } s_{X} < -0.20 \vspace{0.2cm} \\ 
    \overline{J}_{X} - \mathrm{\overline{M}}_{J_{SMC}},& \text{otherwise}
\end{cases}
\label{eq:dismod}
\end{equation}
where the value $-0.20$ is approximately midway between the skewness of the two galaxies.  \removed{Alternatively we could look at the difference between the mode and the median of the distributions which would lead to similar conclusions.} In the above equations, $\mathrm{\overline{M}}_{J_{LMC}}$ and $\mathrm{\overline{M}}_{J_{SMC}}$ are set by adopting  $\mu_{\mathrm{LMC}}$ = 18.477 $\pm$ 0.004 (statistical) $\pm$ 0.026 (systematic) for the LMC \citep{2019Natur.567..200P} , and $\mu_{\mathrm{SMC}}$ = 18.96 $\pm$ 0.03 (statistical) $\pm$ 0.05 (systematic)  for the SMC \citep{2016ApJ...816...49S} .

\inserted{To determine the distance modulus, the median magnitude is preferred over the mode of the LF. This is because for a given sample of CS, the statistical errors are larger on the mode than on the median. Even though the mode is not used for estimating the distance, the fitting procedure is still necessary to obtain the skew parameter and classify the target galaxy.}


\subsection{NGC 6822}

Looking at Fig. \ref{fig:lumfuncs} we notice immediately that the CS luminosity function of NGC 6822 is strongly skewed towards brighter magnitudes. If we compare the best-fit parameters in Tables \ref{tab:best-fit_lmcsmc} and \ref{tab:best-fit_ngcic} we see that, compared to the LMC, the luminosity function of NGC 6822 is only slightly narrower and more skewed towards brighter magnitudes, with moderately less weight on the tails.  \inserted{Not surprisingly the value of the skewness parameter of NGC 6822 is $-0.43$ similiar to the value of the LMC of $-0.44$.} Therefore we use the LMC as the calibrator to calculate the distance to NGC 6822.
Using Eq. \ref{eq:dismod} and the median values obtained from the bootstrapping algorithm (Tables \ref{tab:best-fit_lmcsmc} and \ref{tab:best-fit_ngcic}), we obtain a true distance modulus of $\mu_{0}=23.54\pm0.03$ (statistical). The CS luminosity function of NGC 6822 shifted by $\mu_0$ is shown in Fig. \ref{fig:lumfunc_DM} along with the LMC. 

Table \ref{tab:distngc6822} shows a summary of the distance estimates for NGC 6822  that can be found in the literature. It is important to mention that a direct comparison between the values for $\mu_0$ is not possible as each calculation uses different extinction values and/or zero-points for the calibration of the distance indicator. Nevertheless, our estimate seems to agree better with those that used the same extinction value. NGC 6822 suffers from a substantial amount of extinction, we can see in Table \ref{tab:distngc6822} that the extinction values differ by up to a tenth of magnitude.

\begin{table}
 \caption{Distance estimation to  NGC 6822}
 \centering
 \begin{tabular}{l c c l}
 \hline
    Method & E($B-V$) & $\mu_0^*$ & \multicolumn{1}{c}{Reference} \\
\hline
\hline
   Cepheids & 0.35     & $23.49\pm0.02$ & \cite{2014ApJ...794..107R}\\ Cepheids & (1) & $23.40\pm0.05$ & \cite{2012MNRAS.421.2998F}\\
   Cepheids & 0.26     & $23.49\pm0.03$ & \cite{2009ApJ...693..936M}\\
\hline
   TRGB & 0.35 & $23.54\pm0.05$ & {\cite{2012A&A...548A.129F}}\\ 
   TRGB & 0.24 & $23.34\pm0.12$ & {\cite{2005A&A...429..837C}}\\
\hline
   RR Lyrae & 0.25 & $23.36\pm0.17$ & {\cite{2003ApJ...588L..85C}}\\
  \hline
   Mira & (2) & $23.56\pm0.03$ & {\citealt{2013MNRAS.428.2216W}}\\
 \hline
  C-AGB & (3) & $23.44\pm0.02 $ & {\citet{2020ApJ...899...67F}}\\ 
  C-AGB & 0.35& $23.54\pm0.03 $ & This work \\
\hline
\multicolumn{4}{l}{\small{$^*$ statistical errors only, except for {\cite{2003ApJ...588L..85C}}}}\\
\multicolumn{4}{l}{\small{(1) A$_V=0.667$}}\\
\multicolumn{4}{l}{\small{(2) A$_J=0.20$}}\\
\multicolumn{4}{l}{\small{(3) A$_J=0.167$}}\\ 
 \end{tabular}
 \label{tab:distngc6822}
\end{table}


\subsection{IC 1613}
 We can see in Fig. \ref{fig:lumfuncs} that both the IC 1613 and SMC CS J-magnitude distributions are highly symmetric, as indicated by the low skew parameter. The best-fit parameters summarized in Tables \ref{tab:best-fit_lmcsmc} and \ref{tab:best-fit_ngcic}, indicate that the shape of luminosity function of the CS in IC 1613 is almost identical to that of the SMC. Thus, we categorize IC 1613 as an "SMC-like" galaxy.

\begin{figure*}
\centering
\includegraphics[width=\textwidth]{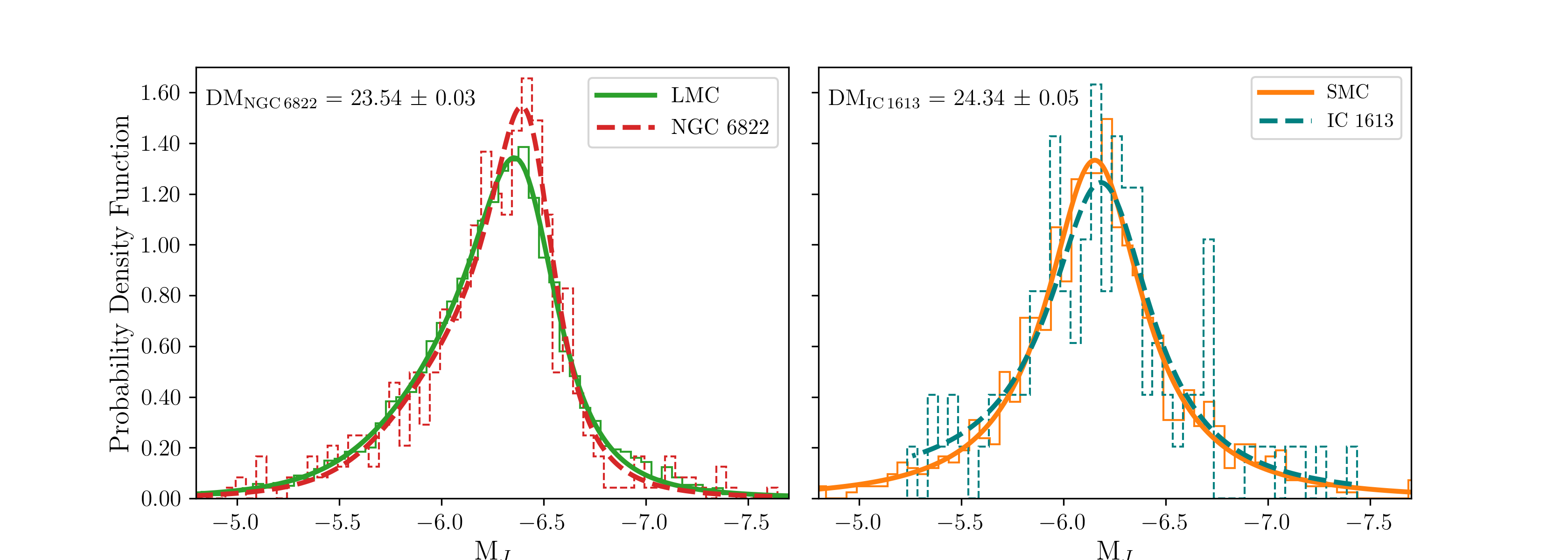}
\caption{CS M$_J$ luminosity functions for LMC (solid green line) and NGC 6822 (dashed red line) in the \textit{left} panel and, SMC (solid orange line) and IC 1613 (dashed dark-cyan line) in the \textit{right} panel. The apparent $J$ magnitude distributions for NGC 6822 and IC 1613 from Fig. \ref{fig:lumfuncs} have been transformed to absolute magnitudes using the distance modulus specified in each plot.}
\label{fig:lumfunc_DM}
\end{figure*}

Using Eq. \ref{eq:dismod} and the median values obtain from the bootstrapping algorithm (Tables \ref{tab:best-fit_lmcsmc} and \ref{tab:best-fit_ngcic}), we obtain a true distance modulus of $\mu_{0}=24.34\pm0.05$ (statistical). Fig. \ref{fig:lumfunc_DM} shows the luminosity functions of the SMC and of IC 1613 shifted by their appropriate distance moduli. 

Table \ref{tab:distic1613} shows a summary of the distance estimates for NGC 6822  that can be found in the literature. As for Table \ref{tab:distngc6822}, the values are taken directly from their sources without any correction for the use of different extinction values and/or zero-points for the calibration of the distance indicator. Unlike NGC 6822, the distance estimations seem to be in much better agreement. Except for \cite{2013MNRAS.435.3206D} RR Lyrae estimation, all the distances in Table \ref{tab:distic1613} agree to within about a tenth of a magnitude.

\begin{table}
 \caption{Distance estimation to IC 1613}
 \centering
 \begin{tabular}{l c c l}
 \hline
    Method & E($B-V$) & $\mu_0$ & \multicolumn{1}{c}{Reference} \\
\hline
\hline
   Cepheids & 0.050 & $24.29\pm0.03$ & {\cite{2013ApJ...773..106S}} \\ 
   Cepheids & 0.025 & $24.34\pm0.03$ & {\cite{2011A&A...531A.134T}} \\
   Cepheids & 0.020 & $24.27\pm0.02$ & {\cite{2009ApJ...695..996F}} \\
\hline
   TRGB & N/A   & $24.30\pm0.05$ & {\cite{2017ApJ...845..146H}}\\ 
   TRGB &  (1)  & $24.39\pm0.03$ & {\cite{2017ApJ...834...78M}}\\
   TRGB & 0.025 & $24.37\pm0.05$ & {\cite{2007ApJ...661..815R}}\\
\hline
   RR Lyrae & N/A & $24.28\pm0.04$ & {\cite{2017ApJ...845..146H}}\\
   RR Lyrae & (2) & $24.19\pm0.09$ & {\cite{2013MNRAS.435.3206D}}\\
 \hline
   Mira & 0.06 & $24.37\pm0.08$ & {\citealt{2015MNRAS.452..910M}}\\
 \hline
   Cep+RR & 0.025 & $24.400\pm0.014$ & {\cite{2010ApJ...712.1259B}}\\
 \hline
   Blue SG & (4) & $24.39\pm0.011$ & {\cite{2018ApJ...860..130B}}\\
 \hline
  C-AGB & (3) & $24.36\pm0.03 $ & {\citet{2020ApJ...899...67F}}\\ 
  C-AGB & 0.022   &  $24.34\pm0.05 $ & This work \\
\hline
\multicolumn{4}{l}{\small{(1) A$_V=0.07$}}\\
\multicolumn{4}{l}{\small{(2) No information given}}\\
\multicolumn{4}{l}{\small{(3) A$_J=0.018$}}\\
\multicolumn{4}{l}{\small{(4) Estimated individually for each star.}}\\
 \end{tabular}
 \label{tab:distic1613}
\end{table}


\section{Three-dimensional structure of the Magellanic Clouds}
\label{sec:3dstruct}

\inserted{A detailed study of the three-dimensional structure of the Magellanic Clouds using CS goes beyond the scope of this paper. Instead, we focus on examining its possible effect in the distance determination method. For this purpose, we divided both the LMC and SMC into nine fields (shown in Fig. \ref{fig:lmc_smc_radec}) and estimated the distance to each of them independently. The fields have equal numbers of CS, with 667-668 CS in each field in the LMC and 93-94 in the SMC. We re-sampled the CS in each field ten-thousand times with replacement. In each iteration we fitted the CS luminosity function with the modified Lorentzian model described in Sec. \ref{subsec:lumfunc} (as we did for the full sample of CS) and obtained the median $J$ magnitude. The parameters obtained for each field from the bootstrapping analysis are shown in Tables \ref{tab:lmc_9fields} and \ref{tab:smc_9fields} for the LMC and SMC respectively. To facilitate the comparison between the results in each field with the parameters determined for the complete CS sample, tables \ref{tab:lmc_9fields} and \ref{tab:smc_9fields} report the absolute median $J$ magnitudes of each field assuming $\mu_{\mathrm{LMC}}$ = 18.477 $\pm$ 0.004 (statistical) $\pm$ 0.026 (systematic) for the LMC \citep{2019Natur.567..200P} , and $\mu_{\mathrm{SMC}}$ = 18.96 $\pm$ 0.03 (statistical) $\pm$ 0.05 (systematic)  for the SMC \citep{2016ApJ...816...49S}. The last row in Tables  \ref{tab:lmc_9fields} and \ref{tab:smc_9fields} detail the difference between the median absolute $J$ magnitude of a specific field and the full CS sample: $\mathrm{\Delta\overline{M}}_{J} = \mathrm{\overline{M}}_{J,\mathrm{F_i}} - \mathrm{\overline{M}}_{J,\mathrm{all}}$. The difference in distance modulus of each field with respect of the entire LMC or SMC is reflected in (and equal to) $\mathrm{\Delta\overline{M}}_{J}$. }


\subsection{The LMC}

\inserted{Inspecting Table \ref{tab:lmc_9fields} we see that the results for the shape parameters of CS luminosity function fit for each field agree (within the error bars) with those reported for the complete CS samples in Tab. \ref{tab:best-fit_lmcsmc}. Only the skew parameter for F2 could be considered slightly outside of the range determined in Sec. \ref{subsec:method}. }

\inserted{Examining the $\mathrm{\Delta\overline{M}}_{J}$ values for the different fields of the LMC, we notice that for the majority of the fields there are minimal variations with respect to the adopted distance modulus. The exceptions are fields 1, 8 and 9 (F1, F8 and F9), where the divergence of $\mathrm{\Delta\overline{M}}_{J}$ from zero is significant. The median $M_J$ magnitude of the CS in F8 is $0.07$ magnitudes fainter compared to the complete LMC CS sample. F8 partially overlaps with the star forming region 30 Doradus located at $\alpha = 5^h38^m$ and $\delta = -69^\circ06'$ \citep{2000A&A...355L..27H}. 30 Doradus and its surrounding area have been catalogued as the most extincted region in the LMC reaching $E(B-V)$ values of $\sim0.5$ mag \citep{2004ApJ...601..260N, 2011AJ....141..158H, 2016ApJ...832..176I}. The fainter magnitudes in F8 are consistent with the high-reddening estimations for this region. The median $M_J$ for F1 is $0.11$ mag fainter than the complete sample while F9 is $0.12$ mag brighter. This difference in median $M_J$ is equal to the difference between the fields' distance moduli and the distance modulus considered for the entire LMC. Our results show a total variation of $0.23$ mag between the northeast (F9) and southwest (F1) field. This variation suggests that the LMC is tilted along the line of nodes with respect to the plane of the sky, with the northeast field closer to the observer and the southwest field more distant. This is in agreement with multiple studies that have found evidence for and have been able to measure the inclination angle of the LMC, e.g. \cite{1986MNRAS.218..223C, 2001ApJ...548..712W, 2001AJ....122.1807V, 2013A&A...552A.144S, 2018ApJ...858...12H}, etc. As in the present paper, \cite{2001AJ....122.1807V} and \cite{2001ApJ...548..712W} used near-infrared data and AGB stars. Our results are in good agreement with those found by the first authors, who reported magnitude variations of up to $\sim0.25$ mag in $J$, $H$, and $K_s$. On the other hand, \cite{2001ApJ...548..712W} used a smaller number of fields and stars and found a $\sim0.3$ mag gradient in the east-west direction, and no variations in the north-south direction.}

\subsection{The SMC}

\inserted{Inspecting Table \ref{tab:smc_9fields} we see that, as for the LMC, the results for the CS luminosity function best-fit shape parameters for each field in the SMC agree (within the error bars) with those reported for the complete CS samples in Tab. \ref{tab:best-fit_lmcsmc}. Again, we have one field, F8, whose skew parameter could be considered an outlier.}

\inserted{In the case of the SMC, the $\mathrm{\Delta\overline{M}}_{J}$ values do not show any significant variations from zero except for F4. The SMC structure is known to be more complex  and have a significant line of sight (LOS) depth compared to the LMC. Studies have found LOS variations that go from a few kiloparsecs \citep{2011MNRAS.415.1366K, 2012AJ....144..107H, 2017MNRAS.472..808R, 2018MNRAS.473.3131M} to more than 20 kpc \citep{2016ApJ...816...49S, 2017MNRAS.472..808R}. Another structural feature is the SMC's elongation from east towards west/southwest with the eastern regions closer to us \citep{2012ApJ...744..128S, 2018MNRAS.473.3131M}. Of the references mentioned, estimates made using RR Lyrae stars yield low inclination angles of only a few degrees with respect to the plane of the sky. In contrast, \cite{2012AJ....144..107H} using Cepheid variables found $i=74^\circ \pm9^\circ$. The estimation of the structural parameters of the SMC seems to depend on the tracers used in and the size of the SMC area covered in each study. \cite{2012AJ....144..107H} with an area comparable to the one used in this study, even though they are able to detect variations in the LOC depth, they claim the field they used is not large enough to deduce the actual shape of the SMC. \cite{2009A&A...496..399S, 2012ApJ...744..128S} used the width of the distribution of magnitudes of red clump stars to estimate the depth of different regions of the SMC. In this study we do not find significant variations in the width of the CS luminosity functions for the different fields. } 


\subsection{Crowding}

\begin{figure}
    \centering
    \includegraphics[width=\columnwidth]{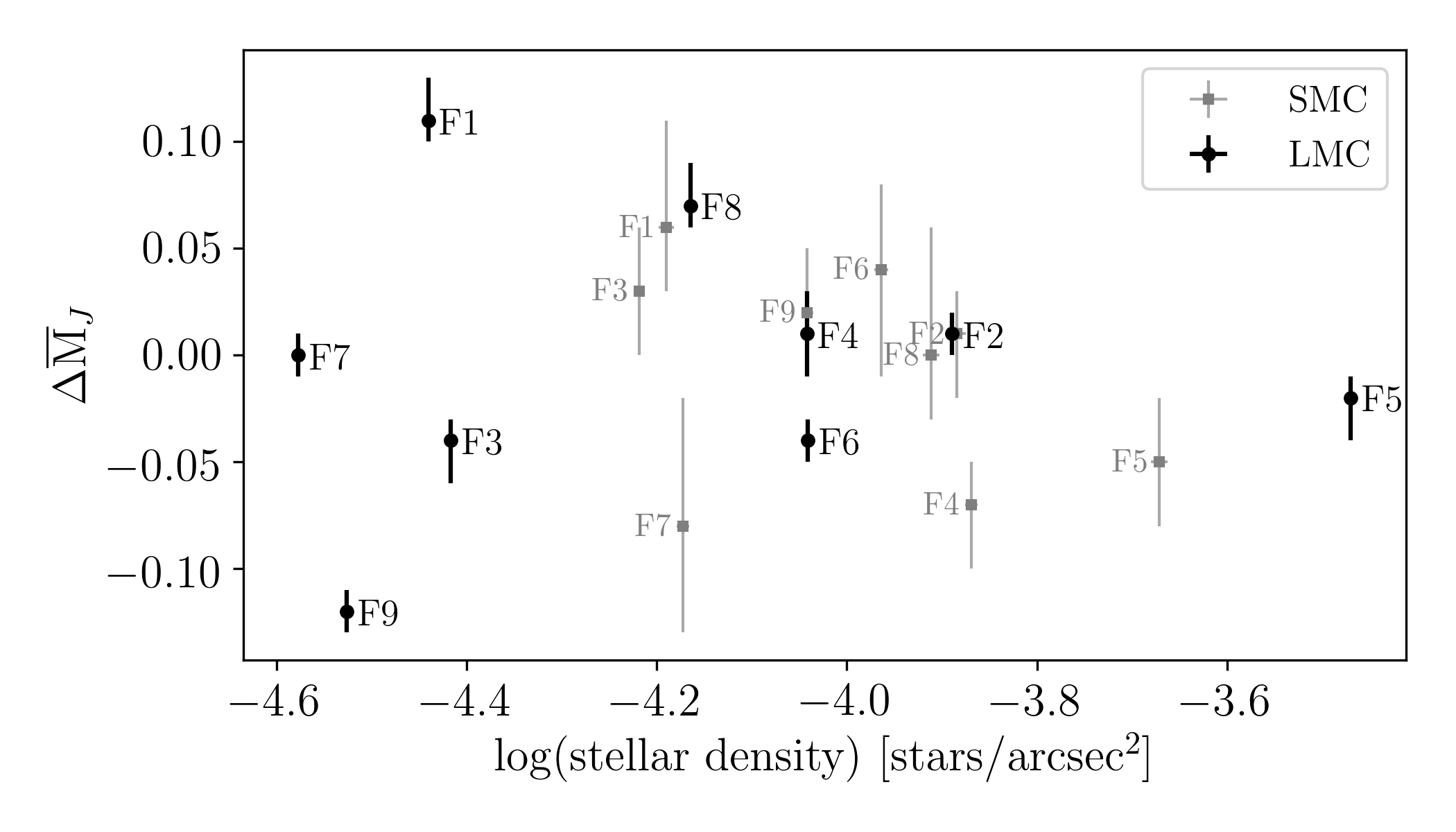}
    \caption{\inserted{Difference between the median absolute $J$ magnitude of a specific field and the full CS sample ($\mathrm{\Delta\overline{M}}_{J})$) versus stellar density. $\mathrm{\Delta\overline{M}}_{J})$ can also be interpreted as the difference in distance modulus. Besides LMC's F1, F8 and F9, that deviate form zero due to intrinsic properties of the galaxy, we see no correlation between the plotted values.}}
    \label{fig:delMvsD}
\end{figure}

\inserted{Our three-dimensional analysis also serves as a test of crowding effects in our results. In both the LMC and SMC, the fields have the same number of stars but cover areas of different dimensions. In Fig. \ref{fig:delMvsD} we plot $\mathrm{\Delta\overline{M}}_{J}$ against stellar density for each field. For the LMC, we have explained that F1, F8 and F9 are outliers. Going through the rest of the fields we can see that stellar density varies up to a factor of $\sim$10 but there is no obvious correlation with $\mathrm{\Delta\overline{M}}_{J}$. In fact, F5 is the most crowded region in both the LMC and SMC and the distances are consistent with $\mathrm{\Delta\overline{M}}_{J}=0$ within the error bars.}

\subsection{Discussion}

\inserted{For both the LMC and SMC, the minimal variations in the luminosity function model parameters tell us that the contribution of CS from different locations in the LMC and SMC does not have a significant impact on the overall estimated luminosity function parameters. Based on the skew parameter, only one field out of nine in each of the Magellanic Clouds could have been paired with the less suitable calibrator. Further research is necessary to determine what produces the slightly different skew values on LMC-F2 and SMC-F8 but it may be a reflection of different star formation histories in different parts of the galaxies. Nevertheless, the inclusion or exclusion of these fields does not alter the classification of the entire galaxy or the estimation of distances to other galaxies. }

\inserted{The variations in $\mathrm{\overline{M}}_{J}$ in F1, F8 and F9  of the LMC with respect to the full sample of CS are explained by the intrinsic three-dimensional structure and properties of the LMC. F1 and F9 follow the orientation of the LMC while F8 is affected by the high extinction in the star forming region 30 Doradus. The variations in three out of the nine fields of the LMC do not affect the total median absolute $J$ magnitude and therefore have no effect in the determination of the distance. The detection of these variations in the distance to different fields serves as a test of the sensitivity of our method.} 

\inserted{In the SMC we covered a smaller-sized total field. In this area of the SMC we do not find evidence of variations in the distance modulus or of differential LOS depth in any of the fields. The inclusion or exclusion of any of these fields does not affect the luminosity function fitting procedure or the distance estimations. It is worth mentioning that in external galaxies the LOS depth is negligible compared to the distance to the galaxy.}

\inserted{When we sum up the stars over an entire galaxy, the signal will always be dominated by the population of stars most representative of the metallicity and star formation history of the galaxy. As we try to estimate distances to more distant galaxies the number of observed CS will decrease and we will want to use every one available. For the most distant galaxies, the stars will be concentrated in one small area and excluding any will reduce the accuracy of the distance estimates. The purpose of this paper is to test the potency of the CS as standard candles. To show that intrinsic properties of the galaxy do not greatly affect the final distance estimation (when using CS) we used the entire sample of CS in our calibrating galaxies.}


\begin{table*}
 \caption{LMC Best-fit parameters, median absolute $J$ magnitude and difference between the median $J$ magnitude for each field outlined in Fig. \ref{fig:lmc_smc_radec} and the entire sample of carbon stars ($\mathrm{\Delta\overline{M}}_{J} = \mathrm{\overline{M}}_{J,\mathrm{F_i}} - \mathrm{\overline{M}}_{J,all}$). $\mathrm{\Delta\overline{M}}_{J}$ can also be interpreted as the difference in distance modulus. As for Table \ref{tab:best-fit_lmcsmc}, values and errors are obtained from 10000 bootstrap realizations. }
 \centering
 \begin{tabular}{l r r r r r r r r r}
 \hline
 & \multicolumn{1}{c}{F1} & \multicolumn{1}{c}{F2} & \multicolumn{1}{c}{F3} & \multicolumn{1}{c}{F4} & \multicolumn{1}{c}{F5} & \multicolumn{1}{c}{F6} & \multicolumn{1}{c}{F7} & \multicolumn{1}{c}{F8} & \multicolumn{1}{c}{F9} \\
 \hline
 \hline
\vspace{0.16cm}
Mode & -6.25$^{+0.03}_{-0.03}$ & -6.31$^{+0.02}_{-0.03}$ & -6.38$^{+0.03}_{-0.03}$ & -6.36$^{+0.02}_{-0.02}$ & -6.39$^{+0.03}_{-0.03}$ & -6.40$^{+0.03}_{-0.03}$ & -6.35$^{+0.02}_{-0.02}$ & -6.32$^{+0.03}_{-0.03}$ & -6.47$^{+0.02}_{-0.02}$    \\
\vspace{0.16cm}
Width & 0.27$^{+0.05}_{-0.04}$ & 0.28$^{+0.04}_{-0.03}$ & 0.28$^{+0.04}_{-0.04}$ & 0.23$^{+0.05}_{-0.04}$ & 0.31$^{+0.05}_{-0.04}$ & 0.30$^{+0.04}_{-0.03}$ & 0.21$^{+0.02}_{-0.02}$ & 0.27$^{+0.03}_{-0.03}$ & 0.19$^{+0.02}_{-0.02}$ \\
\vspace{0.16cm}
Skew & -0.53$^{+0.20}_{-0.26}$ & -0.15$^{+0.12}_{-0.17}$ & -0.39$^{+0.18}_{-0.19}$ & -0.58$^{+0.12}_{-0.15}$ & -0.53$^{+0.17}_{-0.17}$ & -0.39$^{+0.18}_{-0.20}$ & -0.29$^{+0.07}_{-0.08}$ & -0.37$^{+0.09}_{-0.11}$ & -0.44$^{+0.10}_{-0.12}$ \\
\vspace{0.16cm}
Kurtosis & 0.19$^{+0.21}_{-0.09}$ & 0.06$^{+0.08}_{-0.03}$ & 0.11$^{+0.09}_{-0.05}$ & 0.13$^{+0.09}_{-0.05}$ & 0.16$^{+0.11}_{-0.06}$ & 0.12$^{+0.10}_{-0.06}$ & 0.04$^{+0.02}_{-0.02}$ & 0.06$^{+0.04}_{-0.02}$ & 0.08$^{+0.05}_{-0.03}$ \\
\vspace{0.16cm}
Amplitude & 1.47$^{+0.10}_{-0.09}$ & 1.39$^{+0.09}_{-0.08}$ & 1.36$^{+0.10}_{-0.09}$ & 1.48$^{+0.13}_{-0.11}$ & 1.26$^{+0.09}_{-0.08}$ & 1.27$^{+0.08}_{-0.07}$ & 1.67$^{+0.11}_{-0.10}$ & 1.25$^{+0.08}_{-0.08}$ & 1.76$^{+0.12}_{-0.12}$ \\
\vspace{0.16cm}
Median & -6.18$^{+0.02}_{-0.02}$ & -6.27$^{+0.01}_{-0.01}$ & -6.32$^{+0.01}_{-0.02}$ & -6.27$^{+0.02}_{-0.02}$ & -6.30$^{+0.01}_{-0.02}$ & -6.33$^{+0.02}_{-0.01}$ & -6.28$^{+0.02}_{-0.01}$ & -6.21$^{+0.02}_{-0.02}$ & -6.40$^{+0.01}_{-0.01}$  \\
$\mathrm{\Delta\overline{M}}_{J}$ & 0.11$^{+0.02}_{-0.01}$ & 0.01$^{+0.01}_{-0.01}$ & -0.04$^{+0.01}_{-0.02}$ & 0.01$^{+0.02}_{-0.02}$ & -0.02$^{+0.01}_{-0.02}$ & -0.04$^{+0.01}_{-0.01}$ & 0.00$^{+0.01}_{-0.01}$ & 0.07$^{+0.02}_{-0.01}$ & -0.12$^{+0.01}_{-0.01}$ \\
 \hline
 \end{tabular}
 \label{tab:lmc_9fields}
\end{table*}


\begin{table*}
 \caption{Same as Tab. \ref{tab:lmc_9fields} for the SMC.}
 \centering
 \begin{tabular}{l r r r r r r r r r}
 \hline
 & \multicolumn{1}{c}{F1} & \multicolumn{1}{c}{F2} & \multicolumn{1}{c}{F3} & \multicolumn{1}{c}{F4} & \multicolumn{1}{c}{F5} & \multicolumn{1}{c}{F6} & \multicolumn{1}{c}{F7} & \multicolumn{1}{c}{F8} & \multicolumn{1}{c}{F9} \\
 \hline
 \hline
\vspace{0.16cm}
Mode & -6.11$^{+0.12}_{-0.10}$ & -6.13$^{+0.04}_{-0.03}$ & -6.14$^{+0.17}_{-0.07}$ & -6.24$^{+0.06}_{-0.07}$ & -6.16$^{+0.09}_{-0.06}$ & -6.10$^{+0.10}_{-0.09}$ & -6.27$^{+0.10}_{-0.06}$ & -6.24$^{+0.06}_{-0.05}$ & -6.12$^{+0.04}_{-0.04}$    \\
\vspace{0.16cm}
Width & 0.21$^{+0.09}_{-0.10}$ & 0.15$^{+0.08}_{-0.07}$ & 0.20$^{+0.12}_{-0.10}$ & 0.23$^{+0.11}_{-0.08}$ & 0.22$^{+0.08}_{-0.08}$ & 0.26$^{+0.10}_{-0.08}$ & 0.24$^{+0.15}_{-0.13}$ & 0.15$^{+0.08}_{-0.07}$ & 0.15$^{+0.05}_{-0.05}$ \\
\vspace{0.16cm}
Skew & -0.05$^{+0.49}_{-0.48}$ & 0.02$^{+0.24}_{-0.06}$ & 0.00$^{+0.47}_{-0.57}$ & 0.02$^{+0.09}_{-0.50}$ & 0.18$^{+0.33}_{-0.18}$ & 0.05$^{+0.43}_{-0.26}$ & -0.05$^{+0.16}_{-0.24}$ & -0.42$^{+0.28}_{-0.28}$ & 0.16$^{+0.25}_{-0.16}$ \\
\vspace{0.16cm}
Kurtosis & 0.06$^{+0.21}_{-0.06}$ & -0.01$^{+0.04}_{-0.02}$ & 0.04$^{+0.27}_{-0.07}$ &-0.01$^{+0.18}_{-0.04}$ & 0.02$^{+0.10}_{-0.04}$  & 0.03$^{+0.16}_{-0.07}$ & -0.01$^{+0.08}_{-0.04}$ & 0.06$^{+0.13}_{-0.04}$ & 0.02$^{+0.06}_{-0.03}$ \\
\vspace{0.16cm}
Amplitude & 1.57$^{+0.41}_{-0.26}$ & 1.97$^{+0.79}_{-0.45}$ & 1.53$^{+0.53}_{-0.29}$ & 1.42$^{+0.44}_{-0.27}$ & 1.48$^{+0.44}_{-0.25}$ & 1.34$^{+0.30}_{-0.21}$ & 1.26$^{+0.60}_{-0.29}$ & 1.81$^{+0.64}_{-0.34}$ & 2.03$^{+0.54}_{-0.36}$ \\
\vspace{0.16cm}
Median & -6.10$^{+0.06}_{-0.03}$ & -6.15$^{+0.02}_{-0.03}$ & -6.13$^{+0.02}_{-0.03}$ & -6.23$^{+0.01}_{-0.03}$ & -6.22$^{+0.03}_{-0.02}$ & -6.12$^{+0.04}_{-0.06}$ & -6.24$^{+0.07}_{-0.05}$ & -6.16$^{+0.07}_{-0.04}$ & -6.14$^{+0.02}_{-0.03}$  \\
$\mathrm{\Delta\overline{M}}_{J}$ & 0.06$^{+0.05}_{-0.03}$ & 0.01$^{+0.02}_{-0.03}$ & 0.03$^{+0.03}_{-0.03}$ & -0.07$^{+0.02}_{-0.03}$ & -0.05$^{+0.03}_{-0.03}$ & 0.04$^{+0.04}_{-0.05}$ & -0.08$^{+0.06}_{-0.05}$ & 0.00$^{+0.06}_{-0.03}$ & 0.02$^{+0.03}_{-0.03}$ \\
 \hline
 \end{tabular}
 \label{tab:smc_9fields}
\end{table*}

\section{Discussion}
\label{sec:discussion}

\subsection{Using the median vs. the mean}
\label{sec:mVSm}

In contrast with \cite{2020ApJ...899...66M}, we choose to use the median $J$ magnitude instead of the mean. Our selection is based on the fact that the median is less affected by a $J$ magnitude selection; therefore we do not make any magnitude cuts. Instead, we take all the stars in the $1.4 < (J-K_s)_0 < 2.0$ colour range.

\begin{figure}
\centering
\includegraphics[width=\columnwidth]{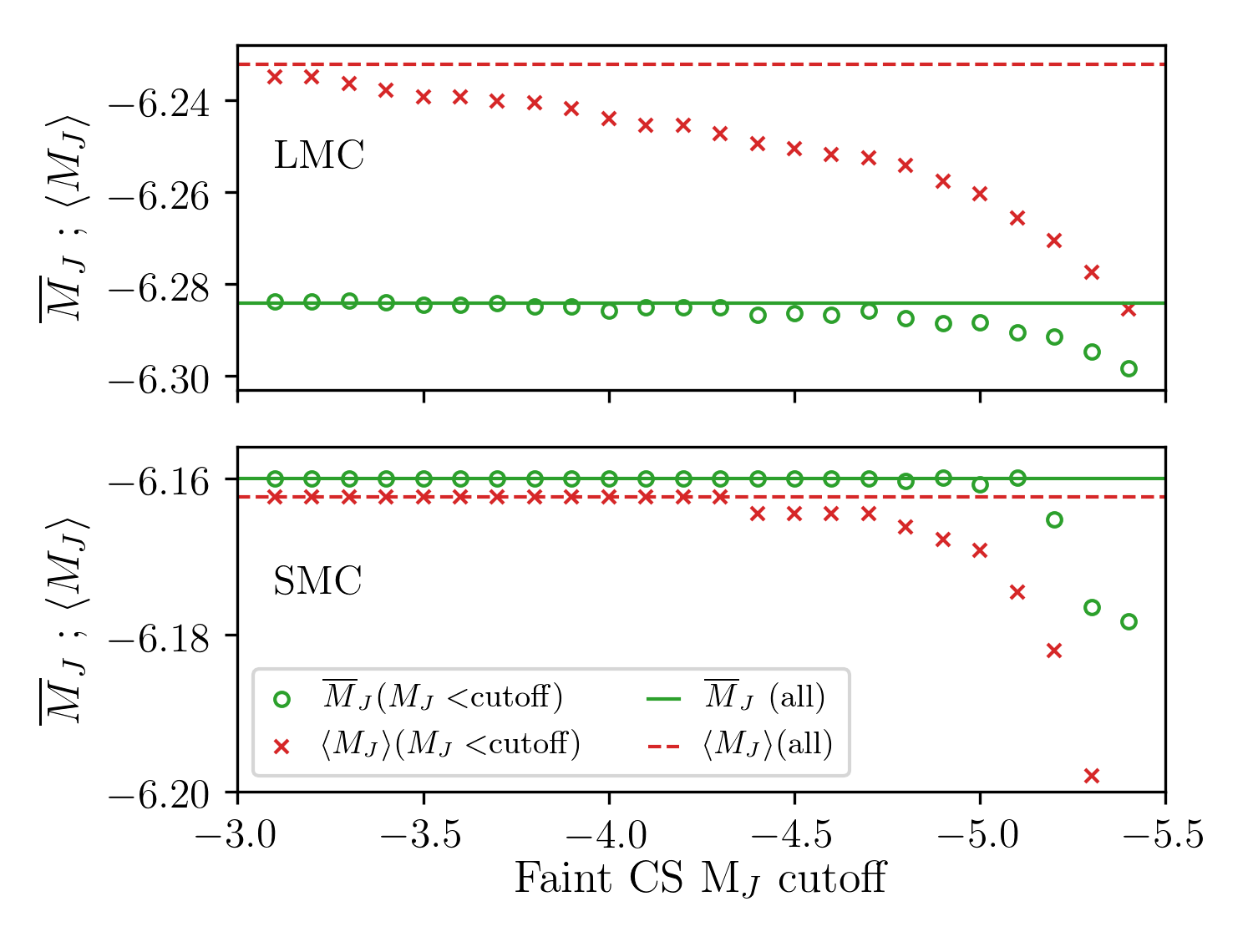}
\caption{Values for the mean ($\left<M_J\right>$; red crosses) and the median ($\overline{\mathrm{M}_J}$; green circles) M$_J$ magnitude for the CS in the LMC \textit{top} and SMC \textit{bottom} using different faint magnitude cutoffs. The magnitude range of the sample goes from $-\infty$ to $\leq$ M$_{J,cutoff}$ The colour selection remains always at $1.4<(J-K_{s})_{0} < 2.0$. The dashed-red and solid-green lines mark the value of the mean and median for the entire sample of carbon stars.}
\label{fig:meanvsmedian}
\end{figure}

For NGC 6822 and IC 1613 it was necessary to establish a faint magnitude limit for the CS samples. We test the effect of a faint magnitude cut by calculating the mean ($\left<M_J\right>$) and the median ($\overline{\mathrm{M}_J}$) of the CS sample choosing different faint magnitude cutoffs for the LMC and SMC. Fig. \ref{fig:meanvsmedian} shows how the mean and the median vary as we push the faint magnitude limit towards brighter magnitudes. We see that for both the LMC and SMC the median varies less than the mean with respect to the values of the full sample. If we look at Fig. \ref{fig:lumfuncs} we see that the luminosity functions start to increase at the faint end at magnitude M$_J=~-5$. The change between the median for all the stars and the median for $-\infty < \mathrm{M}_J \leq -5$ is $0.005$ mag for the LMC and $0.001$ mag for the SMC, while the mean varies $0.025$ mag and $0.01$ mag for the LMC and SMC respectively. The median is not significantly affected by a faint magnitude cut, in fact the variations due to a magnitude cut are comparable to the error in the median. On the other hand, the variations in the mean are of the order of the error bars in the distance modulus obtained for NGC 6822.

If we make the faint magnitude limit for NGC 6822 and IC 1613 fainter or brighter by $0.5$ mag the distance modulus varies by only $\pm~0.005$ mag for NGC 6822 and $\pm~0.05$ mag for IC 1613. The variation in NGC 6822 is lower as we were able to remove most of the contamination to the CMD by identifying them as background galaxies. 

We also note that the selection of CS used by \cite{2020ApJ...899...66M} covers only $\sim0.8$ magnitudes in $J$. If, for the LMC, we consider a selection box centred at M$_J$= -6.2 with $\Delta$M$_J=\pm0.4$ mag we get $\overline{\mathrm{M}_J}=-6.28$ and $\left<M_J\right>=-6.25$. Changing the width of the box by $\pm 0.1$ mag only changes the mean and the median by $\sim \pm 0.01$ mag. Instead, when we change \inserted{the central value of} M$_J$ of the box by $\pm 0.1$ mag (for a box with $\Delta$M$_J=\pm~0.4$), the mean changes by $\pm~0.056$ and the median by $\pm~0.042$. This shows that making a tight magnitude selection has a considerable effect on the value of $\left<M_J\right>$ and $\overline{\mathrm{M}_J}$. We can reduce the impact of choosing at which magnitude to put the box by choosing a wider magnitude range. For example, if we take a box centred at M$_J= -6.2$ but now with $\Delta$M$_J=\pm~1$ mag, when we move the box by $\pm~0.1$ mag the mean changes by $\pm~0.013$ and the median by only $\pm~0.005$.


\subsection{Using the suitable vs. unsuitable Magellanic Cloud}

\begin{figure*}
\centering
\includegraphics[width=\textwidth]{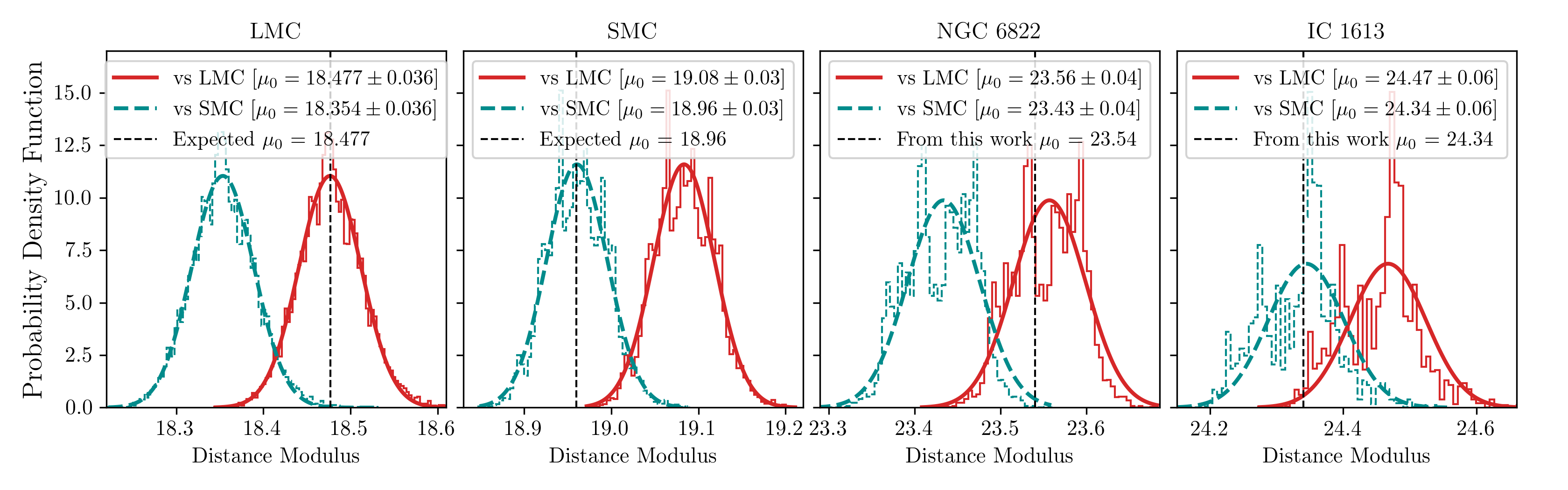}
\caption{\inserted{Distributions of distance moduli (DM) for $10^4$ random subsamples of carbon stars from (from left to right) the LMC, SMC, NGC 6822 and IC 1613. In each plot the red solid line represents the distribution of DM of the specified galaxy using the LMC as calibrator, and the (dashed dark-cyan line) is the DM using the SMC as calibrator. Each carbon star sub-sample is composed of 120 stars, except for the samples from IC 1613 which are composed of 60 stars.} \removed{\textit{Left:} Distributions of distance moduli for the $10^4$ random sub-samples of CS from the LMC using the LMC (solid red line) and SMC (dashed dark-cyan line) as calibrators. \textit{Center:} Same as \textit{left} panel but now the CS sub-samples are drawn from the SMC. \textit{Right:} same as \textit{left} panel but CS are drawn from NGC 6822. Each CS sub-sample is composed of 120 CS.} Measuring the distance modulus to a galaxy using the wrong calibrator galaxy introduces a systematic error of the order of a tenth of magnitude.
}
\label{fig:random_lmcsmc}
\end{figure*}

To test the systematic error introduced by using the less suitable calibrator galaxy we take $10^4$ sub-samples of 120 CS randomly selected from the LMC and calculate the distance modulus using both the LMC and SMC as calibrators. Because the stars in the sub-sample come from the list of CS in the LMC we expect each sub-sample to have the same distance as the LMC. The left panel of Fig. \ref{fig:random_lmcsmc} shows the distributions of distance moduli obtained using the LMC and SMC as calibrators for the $10^4$ random sub-samples. We see that when we measure the distance to the sub-sample using the LMC as calibrator the peak of the distribution of distance moduli is at the expected distance modulus. 

Instead, when we measure the LMC sub-samples against the SMC we get a distance modulus off by $0.123$ mag from the expected value. If we do the same exercise but now with CS drawn from the SMC, we get the expected distance modulus when measuring SMC sub-samples against the SMC and a distance modulus off by $0.12$ mag when SMC sub-samples are measured against the LMC. The distance moduli distribution for the SMC sub-samples are shown in the \removed{right} \inserted{second from the left} panel of Fig.\ref{fig:random_lmcsmc}.\removed{We do a last test } \inserted{We repeat the exercise} using random stars from NGC 6822. When we measure NGC 6822 against the LMC we get a difference of $0.02$ magnitudes with respect to the distance modulus obtained for the full sample, a difference that is well within the statistical errors. Instead, when we use the SMC as calibrator we get a distance modulus that is off by $0.13$ mag. \inserted{For IC 1613 we used smaller sub-samples of only 60 carbon stars. The result for the distance modulus using the SMC is the same as expected from this work. If we use the less suitable calibrator, in this case the LMC, the distance modulus is off by $0.13$ mag. We can also see that the distance estimation to IC 1613 does not change when using only 60\% of the stars. This test on IC 1613 allowed us to show that our method is robust enough that even for a smaller sample of CS the median $J$ magnitude is not affected.}

\section{Conclusions}
\label{sec:conclusion}

We have developed an independent distance determination method using CS as distance indicators with the LMC or SMC as calibrators. It has been tested in two local group galaxies: NGC 6822 and IC 1613. The CS were selected based solely on their colour in the ($(J-K_{s})_0$, $J_0$) CMD. All galaxies were corrected for extinction before selecting the CS. In this way we are able to use the same colour limits for all the galaxies. 

For NGC 6822 we found $\mu_{0,NGC 6822}=23.54\pm0.03$ (stat) adopting $\mu_{0,LMC}=18.477 \pm 0.004$ (stat) $\pm 0.026$ (sys) \citep{2019Natur.567..200P}. Taking into consideration the error bars, the differences in the distance calibration zero-point and the lack of agreement for the extinction, our distance is in good agreement with the distances found in the literature and summarized in Table \ref{tab:distngc6822}. For IC 1613 we obtained $\mu_{0,IC1613}=24.34\pm0.05$ assuming $\mu_{0,SMC}= 18.96 \pm 0.03$ (stat) $\pm 0.05$ (sys). The distance derived here agrees with distances in Table \ref{tab:distic1613}. The error in our distance moduli come from the statistical errors in Tables \ref{tab:best-fit_lmcsmc} and \ref{tab:best-fit_ngcic}. 

The method uses the median $J$ magnitude of the CS and compares it to the appropriate calibrator (LMC or SMC). To determine the most suitable,  calibrating galaxy we first attempted to compare intrinsic properties such as metallicity and SFH. Most studies agree that overall, the LMC and NGC 6822 are more metal rich than the SMC, and that IC 1613 has the lowest metallicity value of the four. Metallicity estimates depend greatly on the tracers used in the analysis and are in general not representative of the CS population. Given the wide spread of values in the metallicity estimates, we find than is preferable to compare the $J-$band luminosity functions of the CS from NGC 6822 and IC 1613 to those from the LMC and SMC. We modeled the CS luminosity functions with a modified Lorentzian distribution (Eq. \ref{eq:lfmodel}) to allow for asymmetry, and used the skew parameter to determine that NGC 6822 is more "LMC-like" and IC 1613 is more "SMC-like". Because in this case the skew parameters are basically identical to the chosen calibrator, we cannot test if the limit defined in Eq. \ref{eq:dismod} is an appropriate limit. We will expand our sample of galaxies in a future publication to include galaxies with a wider range of characteristics and determine whether the skew parameter is the best indicator from which to choose a calibrator. 

We found that constraining the magnitude selection of the CS region considerably impacts the value of the median ($\overline{\mathrm{M}_J})$ and the mean ($\left<M_J\right>$) of the distribution. Choosing a wider range of magnitudes reduces the effects of magnitude selection. The catalogue developed in Paper I for the LMC and SMC allowed us to obtain CMDs clean enough that no magnitude selection was necessary for these data. However for NGC 6822 and IC 1613 the contamination to the CMD at the faint end of the CS region required that we set a faint magnitude limit. Still, the magnitude ranges covered by the CS in these two galaxies was large enough that changing the limit by up to half a magnitude only has an effect on the distance modulus of  the same order as the error in the distance modulus. 

It is important to mention that the method developed here has only been used successfully for Magellanic-type irregular galaxies. As shown in Paper I, the CS luminosity function is notably different for the Milky Way, and we expect this to be the case for all grand design spiral galaxies. In addition, we have shown that choosing the wrong Magellanic Cloud as calibrator introduces an error on the distance modulus of $\sim0.1$ magnitudes. 

\inserted{In the present paper we tested our method in two relatively nearby galaxies. In a future publication we will test this method for galaxies at further distances.}


\section*{Acknowledgements}
This work was supported by the Natural Sciences and Engineering Research Council of Canada, the Canada Foundation for Innovation, the British Columbia Knowledge Development Fund. Part of this research is based on observations obtained with WIRCam, a joint project of CFHT, Taiwan, Korea, Canada, France, at the Canada-France-Hawaii Telescope (CFHT) which is operated by the National Research Council (NRC) of Canada, the Institut National des Sciences de l'Univers of the Centre National de la Recherche Scientifique of France, and the University of Hawaii.

\section*{Data availability}
The LMC and SMC catalogues developed in Paper I can be found at: \url{https://gitlab.com/pripoche/using-carbon-stars-as-standard-candles}. The photometric catalogue for NGC 6822 will be shared upon request to the corresponding author.




\bibliographystyle{mnras}
\bibliography{biblio}





\bsp	
\label{lastpage}
\end{document}